\documentclass[a4paper,11pt]{article}
\usepackage{jheppub}
\usepackage[T1]{fontenc}
\usepackage{axocolor}
\usepackage{cancel}
\usepackage{amsmath}
\usepackage{slashed}
\usepackage{graphicx,caption}
\usepackage{multicol}
\usepackage{footnote}
\usepackage{subfigure}
\usepackage[utf8]{inputenc}
\usepackage{color}
 \usepackage[normalem]{ulem}
\long\def\/*#1*/{}
\usepackage{color}
 \usepackage[normalem]{ulem}
 
\definecolor{darkgreen}{cmyk}{1,0,1,0.4}

\def\barr{\begin{array}}
\def\earr{\end{array}}
\def\dis{\displaystyle}
\def\mev{\, {\rm MeV}}
\def\gev{\, {\rm GeV}}

\def\lapp{\mathrel{\rlap{\raise.5ex\hbox{$<$}}
                    {\lower.5ex\hbox{$\sim$}}}}
\def\gapp{\mathrel{\rlap{\raise.5ex\hbox{$>$}}
                    {\lower.5ex\hbox{$\sim$}}}}
\def\gv{g_{{}_{V}}}
\def\gjv{g_{{}_{VJ}}}

\title{\boldmath Model Independent analysis of MeV scale dark matter:  I. Cosmological constraints}

\abstract{ Recent results from several direct detection experiments
have imposed severe constraints on the multi-GeV mass window for
various dark matter (DM) models. However, many of these experiments
are not sensitive to MeV scale DM as the corresponding recoil
energies are, largely, lower than the detector thresholds. We
reexamine the light scalar DM in a model-independent approach. In
this first of a two-part work, we develop an appropriate methodology to
determine the effective coupling of such a DM to hadrons, thereby
allowing for the determination of the corresponding annihilation
rates. We find that while the parameter space can
be constrained using cosmological and astrophysical observations, a
significantly large fraction is still viable. In the companion
paper, we study the sensitivity of both direct detection experiments
as well as colliders to such a DM.  
\\[1ex] \textbf{Keywords: Dark
    Matter, MeV scale scalar, Belle-II}}

\author[a]{Debajyoti Choudhury,}
\author[a,1]{Divya Sachdeva\note{Corresponding author (divyasachdeva951@gmail.com).}}

\affiliation[a]{Department of Physics and Astrophysics, University of Delhi, Delhi 110 007, India}

\emailAdd{debajyoti.choudhury@gmail.com}
\emailAdd{divyasachdeva951@gmail.com}
\begin{document}
\maketitle
\section{Introduction}
The last half a century has witnessed the accumulation of overwhelming
evidence for gravitational interactions between visible (and stable)
particles and non-luminous matter on a multitude of scales, from the
galactic to the cosmological. Starting with rotation curves in spiral
galaxies~\cite{Rubin:1970zza}, gravitational lensing
measurements~\cite{Moustakas:2002iz,2012A&A...545A..71V}, recent
observations of cluster collisions (Bullet
Cluster)~\cite{Clowe:2006eq}, a temperature anisotropy in the spectrum
of Cosmic Microwave Background
Radiation~\cite{Hinshaw:2012aka,Abadi:2002tt,Ade:2015xua}, there are a
large variety of observations for which the Dark Matter (DM)
hypothesis provides the most compelling explanation.

However, all these observations are indirect and, thus, it is
difficult to ascertain whether it is a particulate DM that underlies
these anomalies or whether the latter are but manifestations of our
lack of understanding of gravity at different cosmological
scales. These issues have, rightly, been explored and several
alternates to standard gravity proposed~\cite{Milgrom:2001ny,Famaey:2011kh}.
On the other hand, it should be noted that even if such modifications
of gravity do exist, theories incorporating these will not be
necessarily unrelated from a model involving particles as DM~\cite{Calmet:2017voc}.

Despite the lack of direct evidence ({\em i.e.}, one achieved under
controlled conditions), it is the notion of particle DM that has seen
the most development. The reasons are twofold. For one, the
experimental discovery of particle DM is more viable in comparison to
the verification of modified gravity models. Moreover, DM candidates
arise naturally in a host of particle physics models that are
motivated primarily to address other issues that are unanswered by the
Standard Model (SM).  To substantiate such models, a variety of
experiments have been proposed, with many being already under
operation (or even having outlived their use). In principle, this
could be done in three ways: \textit{a)} Satellite based indirect
detection experiments like Fermi-LAT~\cite{Ackermann:2015zua},
PAMELA~\cite{Adriani:2008zr}, AMS~\cite{Lin:2015taa}, depend on the
annihilation of a pair of DM particles into SM particles which can
produce rare antimatter cosmic rays (positrons, anti-protons or
antideuterons), neutrinos, monochromatic photons or continuous
$\gamma$-ray spectrum. Although there, occasionally, have been claims
of anomalies in the data, unfortunately the experiments have failed to
validate each other's putative positive sightings, resulting in
further constraints.  \textit{b)} Direct detection experiments that,
typically, identify the nuclear recoils produced by the scattering
between DM and the detector's (target) nuclei. \textit{c)} Collider
searches based on the production of DM look for the excesses (over the
SM expectations) in final states with large missing momentum. (Note,
though, that collider experiments are indicative at best, for these
can only verify the stability of the putative DM candidate over
detector dimensions, not over cosmological timescales.)

Taking cue from recent null results in the LUX~\cite{Akerib:2016vxi},
PandaX-II~\cite{Tan:2016zwf} and XENON100~\cite{Aprile:2016swn}
experiments for $m_{DM}>$ 6 GeV, we concentrate, here, on MeV scale
DM($m_{DM}\lapp 3 \gev$).  Light DM can easily evade many of these
direct and indirect detection experiments because of the low momentum
transfer (lower than the threshold)\footnote{However, if the DM family
  has light as well as heavy DM, then it can be detected in (direct)
  detections as in the case of boosted DM. In this case, DM can be
  energetic if heavy DM decays into the lighter one. Such DM
  candidates are not included in our analysis, as the details of the
  particular model are paramount in such scenarios, whereas we attempt
  only a model-independent analysis.}, and, consequently, the lack of
signals in this range has motivated discussions of a DM with a
particularly low mass~\cite{Alexander:2016aln}. Similarly, to explain
perceived anomalies in the 511 keV $\gamma$-rays observed by INTEGRAL,
the cosmic $\gamma$-ray background at 1-20 MeV and the details of
large scale structure, quite a few
models~\cite{Boehm:2003bt,Ahn:2005ck,Borodatchenkova:2005ct,Hooper:2007tu}
with a light DM were invoked. Again, WIMPless DM also accommodates DM
masses in the MeV scale~\cite{Feng:2008ya}.  These models emerge
naturally from gauge-mediated supersymmetry breaking where DM
naturally satisfies the current relic density without its mass and
interaction being restricted to the weak scale.

All of the above mentioned models are well motivated but lack
experimental support. Moreover, in case the DM particle is the only new
particle (in the dark sector) within the reach of a particular
experiment while other new species are much heavier, it will be very
difficult to distinguish the underlying theories. Therefore,
model-independent studies of the DM are powerful as they proffer a way
ahead without being constrained to a very specific scenario.

The purpose of this article is to study the parameter space for MeV
scale DM in a model-independent way. We first construct effective
operators describing interactions between a scalar DM particle and the
(visible) SM sector. The constraints on these, as obtained from DM
relic abundance, CMB and the counting of relativistic degrees of
freedom are then analysed. 

\section{Higher Dimension operators}
\label{operators}
The interaction of the dark matter with the SM sector is completely
unknown barring, of course, the gravitational one. All that we know is
that the DM does not interact either strongly or
electromagnetically\footnote{While models have been proposed wherein
  the DM does have a very tiny charge, these tend to be baroque, and
  do not fit within well-motivated scenarios going beyond the
  SM\cite{Overduin:2004sz}.}.  On the other hand, if the DM does not
interact at all with the SM particles (except gravitationally), then
there would, essentially, be no way to directly confirm their
existence. More importantly, with the DM particles having been
produced profusely during the post-inflation reheating phase and
shortly thereafter, without such interactions, the relic density today
would tend to be too large, thereby more than over-closing the universe, an
eventuality that can be avoided only by tuning the initial conditions.

The interaction that, thus, must be posited could be in the form of a
detailed and ultraviolet-complete model (such as that in the minimum
supersymmetric standard model) or in the shape of an effective field
theory. It is the latter approach that we adopt here, choosing to
profess an ignorance of the underlying theory (the UV-completion). In
other words, we would augment the SM with the DM particle and posit
that the latter interacts with the known particles through certain
higher-dimensional operators (without any explicit mediator being
considered). With the typical energy scale of the processes under
consideration being much smaller than the dominant mass scale of the
theory, such an approach is irreproachable.

Our assumption, thus, is that the only new relevant field is the
scalar\footnote{The results are identical for a pseudoscalar DM.},
with all other new species being too heavy to be relevant 
in the contexts of both terrestrial experiments/observations
as well as the cosmological evolution of the
relic density. Since we are interested in a DM with a mass of at most
a few GeVs, the only relevant SM states are the photon and the gluon,
the leptons (including neutrinos) and the quarks of the first two
generations. The bottom-quark is, at best, only marginally
relevant. Consequently, we consider operators including 
this limited set of particles alone. Furthermore, to be
consistent with low-energy constraints, we do not admit flavour
changing operators.
 
Assuming $SU(3) \otimes U(1)_{\rm em}$ symmetry\footnote{All but
  ${\cal O}^f_{s,p}$ respect the full SM gauge symmetry. These two too
  can be altered trivially to respect the full symmetry, albeit at the
  cost of introducing an extra factor of $\langle H\rangle / \Lambda$,
  where $H$ is the SM Higgs doublet. For the present analysis, this distinction 
  is essentially irrelevant.}, the operators for a
complex scalar field\footnote{It should be noted here that analogous
  results can be achieved for a real scalar as well.}  are
\begin{equation}
\begin{array}{lcl}
\mathcal{O}^f_{s}&=& \dis \frac{\mathcal{C}^f_{s}}{\Lambda} \varphi^\dagger\varphi
   \; \bar{f} f \\[2ex]
\mathcal{O}^f_{p}& = &\dis \frac{\mathcal{C}^f_{p}}{\Lambda} \varphi^\dagger\varphi \,
         \bar{f}\gamma^5 f
\\[2ex]
\mathcal{O}^f_{v}& = &\dis \frac{\mathcal{C}^f_{v}}{\Lambda^2}i~( \varphi^\dagger\partial_{\mu}\varphi-\partial_{\mu}\varphi^\dagger \varphi) \; \bar{f}\gamma^{\mu} f 
\\[2ex]
\mathcal{O}^f_{a}& = &\dis \frac{\mathcal{C}^f_{a}}{\Lambda^2}i~( \varphi^\dagger\partial_{\mu}\varphi-\partial_{\mu}\varphi^\dagger \varphi) \; \bar{f}\gamma^{\mu}\gamma^5 f
\\[2ex]
\mathcal{O}_{\gamma}& = &\dis \frac{\mathcal{C}_{\gamma}}{\Lambda^2}( \varphi^\dagger \varphi) F_{\mu\nu} F^{\mu\nu}	
\\[2ex]
\mathcal{O}_{\tilde{\gamma}}& = &\dis \frac{\mathcal{C}_{\tilde{\gamma}}}{\Lambda^2}( \varphi^\dagger \varphi) F_{\mu\nu} \tilde{F}^{\mu\nu} \ ,	
\end{array}
\label{the_operators}
\end{equation}
where $f$ is an arbitrary SM fermion and $\Lambda$ is the scale of new
physics.  Note that we could also write operators akin to
$\mathcal{O}_{\gamma}$ and $\mathcal{O}_{\tilde{\gamma}}$, but for the
gluons instead.  As for the ${\cal C}$'s (the dimensionless Wilson
coefficients corresponding to the various operators), we would be
normalizing these to either zero or unity (denoting the absence or
presence of the said operator).  The results will, thus, depend on the
mass of DM and the the scale $\Lambda$.  Indeed, with each operator
presumably arising from a specific DM-SM interaction in a UV-complete
theory, the conclusions reached from an analysis such as ours can be
easily rescaled to obtain constraints on the parameters of the
underlying theory.

\section{Relic Abundance} 

The model independent framework developed in the preceding section
allows us to constrain the parameter space for any MeV-scale
spin-0 Dark Matter candidate. Particular
attention needs to be paid to the constraints from the relic
abundance, the cosmic microwave background and, on account of the
lightness, that from the counting of relativistic degrees of
freedom. In this section, we consider only the first, relegating 
the others to a later section.

\subsection{The formalism} 
We restrict our discussions to the context of DM that had, primarily,
been produced thermally and was in equilibrium with the SM sector. The
relic abundance of non-thermal DM, on the other hand, depends
crucially on the conditions when it was produced. This, being
intricately tied to the specifics of the dynamics can only be
addressed within the context of a particular model, and, hence, does
not fall under the ambit of a model-independent analysis such as ours.
Since the WIMPs are presumed to be produced thermally, the relic
abundance calculation\cite{Steigman:2012nb,Yu:2011by} can proceed as
usual while taking care of some subtleties owing to the small mass.
For the WIMP to stay in thermal equilibrium, it needs to
interact with the SM sector, with the strength(s) being sufficiently
large to beat the expansion rate of the universe. 

During its evolution, the (stable) spin-0
particle $\varphi$ (of mass $m_{\varphi}$) was in thermal equilibrium until
a certain epoch. Similarly, the SM
particles (barring, possibly, the neutrinos) 
were also in thermal equilibrium with the photon gas. The
latter determines the temperature of the thermal soup, and this we
shall denote by $T_{\gamma}$. The evolution of $\varphi$ is given by the
Boltzmann equation, namely
    \begin{equation}
    \frac{dn}{dt} ~ + ~ 3H(t)n = -\langle\sigma v\rangle (n^2-n_{eq}^2) \ ,
       \label{eq_Boltz}
    \end{equation}
where $n$ is the number density of $\varphi$ ($n_{eq}$ being its
  equilibrium value), $H$ is the Hubble expansion rate and $\langle \sigma
v\rangle$ is the thermally averaged cross-section for DM annihilation.

Before attempting a general solution of eq.(\ref{eq_Boltz}), let us
consider some general properties. For any massive particle, the number
density at equilibrium depends on the ratio of its mass and the
temperature of the plasma, $x \equiv m_{\varphi}/T$.  For a stable
particle that is relativistic yet at equilibrium ($x\ll 1$), the
annihilation processes as well as pair production are proceeding at
comparable rates. The consequent equilibrium density is given by
$n=3\zeta(3)g_{\varphi}T^3/(4\pi^2)$ where $g_{\varphi}$ denotes its degrees
of freedom. On the other hand, if it is nonrelativistic, {\it i.e.},
its mass is much larger than the ambient energy ($x\gg 1$), the plasma
does not have sufficient energy to drive pair production; yet, the
pair annihilation proceeds. Consequently, its equilibrium abundance
falls exponentially as the temperature drops below the mass of the
particle, yielding
$n=g_{\varphi}(m_{\varphi}T/(2\pi))^{3/2}e^{(-m_{\varphi}/T)}$.

Applying the above to the WIMP, in the immediate aftermath of its
production, it would have been in equilibrium due to the balance
between its interactions with the SM particles as its interactions
with this sector were strong enough to beat the expansion rate. If it
continues to be in equilibrium, then as the universe cools to $T \ll
m_\varphi$, the WIMP would become nonrelativistic and its abundance today
would have been negligible. However the very structure of its
interactions (essentially the higher-dimensional nature of the
couplings), stipulates that it must fall out of equilibrium for $T \ll
\Lambda$.  Naively, it might seem that this would have occurred when
the WIMP was still ultrarelativistic, since $m_\varphi \ll \Lambda$.  On
the other hand, N-body simulations\cite{Abadi:2002tt} for structure
formation requires the DM to be non-relativistic\footnote{In this
  scenario, DM perturbations grew in the matter dominated era forming
  a gravitational well. (This could not have been initiated by
  ordinary matter as it could not have clustered due to the radiation
  pressure.)  Ordinary matter could now fall into this well, thereby
  allowing an early start of the structure formation and formation of
  the fine structure in universe.}. In other words, $\varphi$ should
decouple from the thermal soup only when it had become
non-relativistic in the radiation dominated era. This could happen if,
thanks to the exponential suppression of $n_{eq}$, the WIMPs became so
rare that the interaction rate fell below the expansion rate. No
longer affected by interactions, these fall out of equilibrium with
the abundance freezing out, ({\em i.e}, their number in a comoving
volume becomes constant).
 
The freeze-out temperature $T_f$, namely that at the epoch when the DM
number density freezes out, can be determined in terms of the mass and
the interaction strengths, as we discuss now. In the
radiation-dominated regime, it is useful to express the total energy
density in terms of the photon energy density thereby defining
$g_\rho$, the effective number of degrees of freedom associated with
total energy density, namely
\[
g_\rho=\sum_{\rm bosons} g_{\rm bosons}\left(\frac{T_{\rm bosons}}{T_{\gamma}}\right)^4 + \frac{7}{8}
\sum_{\rm fermions} g_{\rm fermions} \left(\frac{T_{\rm fermions}}{T_{\gamma}}\right)^4 \ .
\]
In addition to this, the entropy($S$) in a comoving volume ($S=s
a^3$) is a conserved quantity which enables us to define the number
density in terms of the ``yield'' $Y=n/s$ and
also, analogously, introduce $g_s$, the number of
  effective degrees of freedom associated with the entropy:
\[
g_s=\sum_{\rm bosons} g_{\rm bosons}\left(\frac{T_{\rm bosons}}{T_{\gamma}}\right)^3 + \frac{7}{8}
\sum_{\rm fermions} g_{\rm fermions} \left(\frac{T_{\rm fermions}}{T_{\gamma}}\right)^3\ .
\]

With the density of a nonrelativistic species falling faster, 
$g_\rho$ and $g_s$ differ noticeably only when
there are relativistic particles present that are not in equilibrium
with photons. Within the SM, this occurs for
neutrinos. Similarly, for $m_{\varphi}\lapp 6$ MeV, the DM
will contribute to relativistic degrees of freedom and, thence,
entropy. For such light DM, the explanation of the relic abundance by
way of the DM having reached thermal equilibrium with the SM sector is
excluded by current observations which we discuss in the
next subsection. Although there are models which use a different line
of approach, namely asymmetric or non-thermal~\cite{Iminniyaz:2011yp,Dev:2013yza},
we shall desist from doing so, and will no longer
consider this range. Therefore, we can 
safely assume $g_\rho \simeq g_s$ in our analysis.

Entropy conservation also implies $a(t)T=$ constant (here,
  $a(t)$ is the scale factor of the universe) and, hence,
$dT/dt=-H(t)T$.  Effecting a change of variables, $T \to x \equiv
m_{\varphi}/T$, we, then, have the famous Boltzmann equation, {\it viz.}
\begin{equation}
\frac{dY}{dx} = \frac{m^3_{\varphi} \, \langle\sigma \,v\rangle}{H(m_{\varphi}) \,
x^{2}}\,
                 \left( Y_{eq}^{2} - Y^{2}\right) \ ,
\label{eqn:boltzeqna}
\end{equation}
where $H(m_\varphi)$ is the Hubble expansion rate
(in the radiation dominated universe) calculated at the
  epoch when the temperature equals $m_\varphi$ and is given 
by\footnote{In the radiation dominated universe, the 
scale factor $a(t)$ goes as $t^{1/2}$, while the 
temperature-time relation is given by $ t^2 \, T^4 = 45/8\pi^3 G$. Together,
this gives $H(T)$.}
\[
H(T =  m_{\varphi})= \sqrt{\frac{4 \, \pi^3 \, G \, 
g_{\rho}}{45}} \, m^2_{\varphi} \ .
\]

A caveat needs to be entered here. A key ingredient in reaching
eq.(\ref{eqn:boltzeqna}) is the assumption that entropy is conserved
throughout the era of interest. However, this statement may not
necessarily be true for MeV-range dark matter, especially if
$500\mev\leq m_{\varphi} \leq 1\gev$ (see Fig.~\ref{fig:xf}). The freeze
out temperature (equivalently, $x_f \equiv m_{\varphi}/T_f$) for this
range may lie around the QCD phase transition\footnote{See
  Sec.\ref{sec:relativistic_dof} for a derivation of $x_f$.}.
Consequently, the entropy may not be conserved at this
epoch\footnote{It has been suggested~\cite{Aoki:2006we}, though, that
  the QCD transition in the early universe is not a real phase
  transition but an analytic cross-over. As we argue next, the distinction is of 
  no consequence in the current context.}. However, as the entropy is
overwhelmingly determined by the contributions from relativistic
particles such as $e^{\pm}$, $\gamma$ and $\nu$'s, this small possible
non-conservation can be neglected altogether.

To obtain the present
density of DM particles, we need to 
solve eq.(\ref{eqn:boltzeqna}) in terms of the final freeze out
abundance $Y_{\infty}$ (at $ x = \infty$). While this, unfortunately,
can be done only numerically, it is instructive to consider an
approximate analytic solution. Before freeze-out, $Y_{\varphi}$ was close
to its equilibrium value, $Y_{eq}$
\[Y_{eq}= \dis \frac{45}{2\pi^4} \, \sqrt{\frac{\pi}{8}} \, \frac{g_\varphi}{g_\rho} \, x^{3/2} \, e^{-x}
\]
which is exponentially suppressed\footnote{Although,
    after freeze out, the abundance is larger than what its
    equilibrium value would have been, this approximation of $Y
    \approx Y_{eq}^{\rm freeze-out}$ is an excellent one, especially
    for understanding the structure of the
    solution.}. 
Integrating eq.(\ref{eqn:boltzeqna}) from the freeze out 
temperature  $x_{f}$ until very late times ($x = \infty$),
we get 
\[
Y_{\infty} \simeq x_f \, H(m_{\varphi})/m^3_{\varphi} \langle\sigma v\rangle \ .
\]
With the energy density for the now non-relativistic DM, being given by
$\rho_{\varphi} = m_{\varphi} \, n_{\varphi}$, post freeze-out, it simply
falls as $a^{-3}$. Denoting
 the freeze-out epoch ({\it i.e.},
when $Y$ has reached the asymptotic value $Y_{\infty}$) by the
temperature $T_f$ and the scale factor $a_f$, with the corresponding
quantities today being given by $T_0$ and $a_0$ respectively,
we have $n(a_f,T_f) = Y_{\infty} \, T^3_f$,
and, today, 
\[
\rho_{\varphi} = m_{\varphi} Y_{\infty} T_0
^3 \left(\frac{a_f T_f}{a_0 T_0}\right)^3 \ .
\] 
Simultaneously, the 
number of effective degrees of freedom changes from  $g_{\rho}(x_f)$ 
at the freeze out epoch 
to $g_0 = 3.36$ operative today, and $g_0 \, a_{0} \, T^3_{0} = g_{\rho}(x_f)
\, a_{f} \, T^3_{f}$.
It is customary to parametrize
$\rho_{\varphi} \equiv \Omega_{\varphi} h^2 \, \rho_c $, 
where $\rho_{c}=1.05375 \times 10^{-5} \, h^2 \left({\rm GeV}/c^2 \right) {\rm cm}^{-3}$
is the critical density of the universe with
the Hubble constant 
today being expressed 
as $H_0 = h \times 100$~km\,s$^{-1}$\,Mpc$^{-1}$. We have, then,
\begin{equation}
\Omega_{\varphi} h^2 = \sqrt{\frac{4 \, \pi^3 G g_{\rho}(x_f)}{45}} \; 
\frac{x_f \, T^3_0 \,  g_0}{\rho_c \, \langle\sigma \, v\rangle \,
g_{\rho}(x_f)}
\ .
\label{omega}
\end{equation}  

Both the WMAP~\cite{2013ApJS..208...19H} and the
Planck~\cite{Ade:2015xua} satellite observations determine the relic
density very well, with the latter suggesting $\Omega_{DM} h^2 =
0.1199 \pm 0.0022$. Clearly, we must have
$\Omega_{\varphi}\leq\Omega_{DM}$. This can be translated to constraints
on the parameter space available to the theory. As the interaction
between the DM and the SM sector increases, so does $\langle\sigma \,
v\rangle$, and, as a consequence, the relic abundance of DM today
decreases (see eq.~\ref{omega}).

\subsection{Bounds from Relic abundance}

The dimension-5 operators in eq.(\ref{the_operators}) would lead to
$\sigma \propto \Lambda^{-2}$. On the contrary, for nonrelativistic particle,
the dimension-6 operators stipulate $\sigma \propto m_\varphi^2 \,
\Lambda^{-4}$. Consequently, for the second set, a given value of
$\Omega_{\varphi}h^2$ would require $m_\varphi$ to increase nearly
quadratically with $\Lambda$. On the other hand, for the first set,
the ``right value'' of $m_\varphi$ should have only a weak dependence on
$\Lambda$.
   
\begin{figure}[htb]
\centering
 \subfigure[]{\label{r1}\includegraphics[width=75mm]{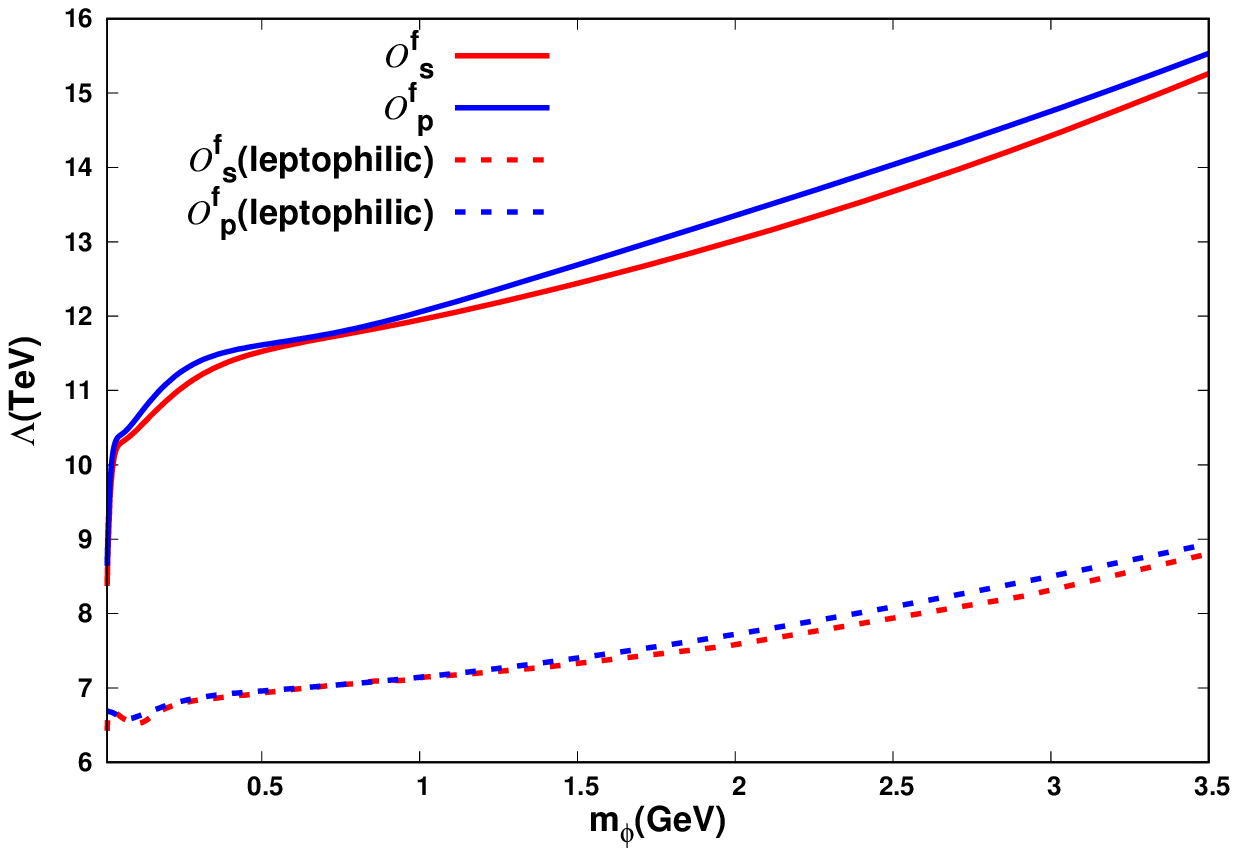}}
  \subfigure[]{\label{r2}\includegraphics[width=75mm]{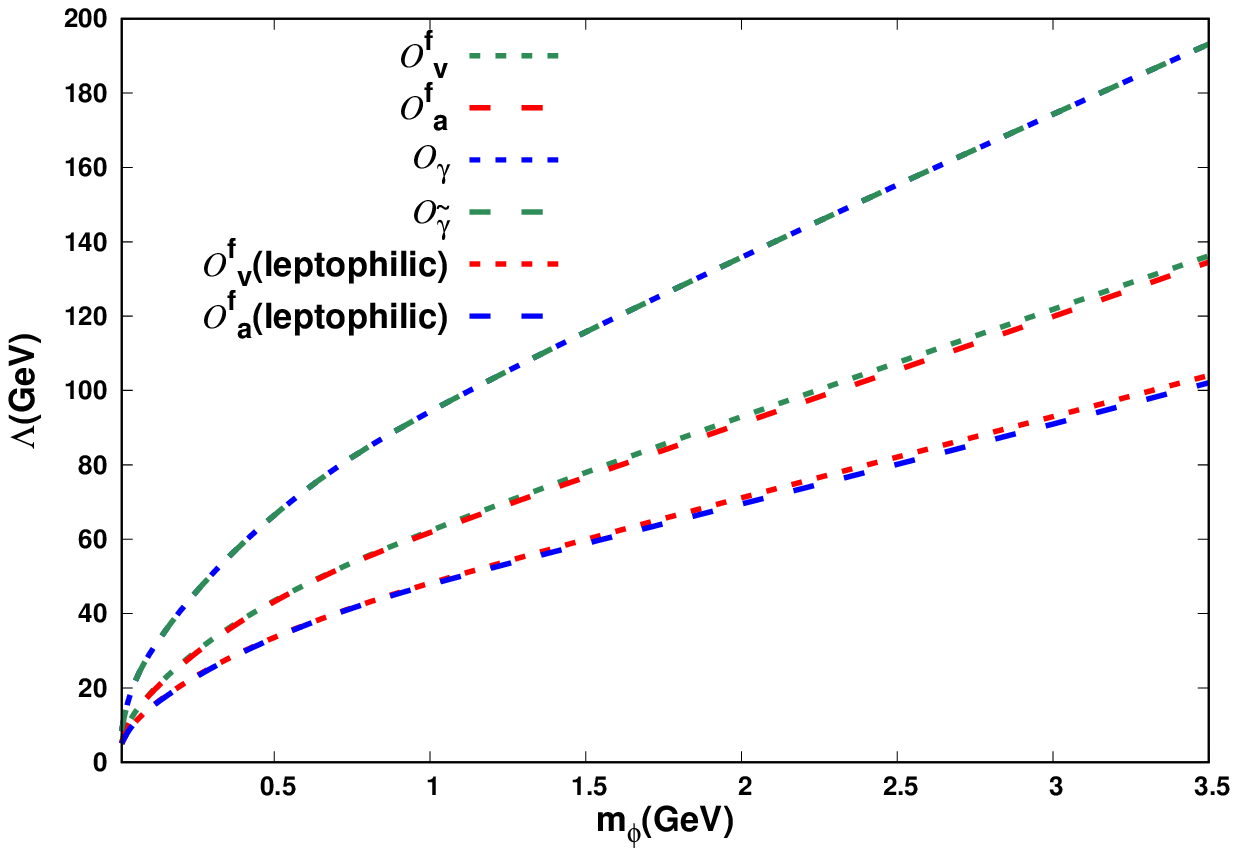}}
  \caption{\em Contours in the $m_{\varphi}-\Lambda$ plane for {\em
      (a)} the dimension-5 operators in eq.(\ref{the_operators}) and
    {\em (b)} the dimension-6 operators in eq.(\ref{the_operators}),
    satisfying $\Omega_{DM} h^2 = 0.1199 \pm 0.0022$. The error bars
    are subsumed in the thickness of the curves. Only one of the new
    physics operator structures is deemed effective. For the fermionic
    operators, all the SM fermions (leptons) participate
    equally---with ${\cal C}_\alpha^f = 1$--- in the general
    (leptophilic) case. Wherever applicable, open quark production has
    been considered, postponing consideration of bound state effects 
    until later (see text).}
\label{fig:relicdensity} 
\end{figure}

To delineate the exact parameter space (as opposed to basing our
conclusions on analytic results obtained from an approximations as we
have been making so far), we implement these additional operators in
micrOMEGAs~4.1\cite{Belanger:2014vza} using
FeynRules\cite{Alloul:2013bka}.  To best understand the consequences,
we only incorporate a single operator structure (from
eq.~\ref{the_operators}) at a time. For the operators involving a
fermionic current, we consider two different cases, namely $(a)$ one
wherein all the SM fermions participate equally\footnote{Clearly,
  given the mass range of the scalar, the third-generation quarks play
  little or no role.}, i.e., all the ${\cal C}^f$s of a given class
are unity and $(b)$ the leptophilic case, namely where only the
leptons participate (and equally) while the quarks do not. In
Fig.\ref{fig:relicdensity}, we depict the contours in the
$m_\varphi$--$\Lambda$ plane corresponding to $\Omega_{\varphi} h^2 =
0.1199$. The width in the contour due to the measurement error
  in $\Omega_{\varphi}h^2$ is virtually unobservable. In each case, the
area below the curves would correspond to a larger annihilation
cross-section (thanks to a smaller $\Lambda$) and, hence, a DM relic
density smaller than what the Planck collaboration measures. In other
words, this is the parameter space that is observationally allowed
(with the remainder ostensibly being contributed by some other
source). Note that, for the dimension-6 operators (Fig.\ref{r2}), the relation between $m_\varphi$ and
$\Lambda$ is nearly quadratic, as expected. On the other hand, for the
dimension-5 operators (Fig.\ref{r1}),
$\Lambda$ increases much slower with $m_\varphi$. Understandably, the
dimension-5 operators are sensitive to much larger values of
$\Lambda$.

Expectedly, for the leptophilic DM, the value of $\Lambda$ for a given
$m_\varphi$ is much lower than the case of the democratic
coupling\footnote{Here, we assume the simplistic point of view that
  the light quarks can be treated as (pseudo-)asymptotic states. This,
  of course, is untenable, and has been assumed only for illustrative
  purposes. We return to this point later.}. This is but a reflection
of the fact that, now, fewer final states are available to the DM
annihilation process. And, finally, the dependence on the chirality
structure is rather minimal, a reflection of the fact that the fermion
masses in the problem are quite small.

The corresponding constraints for a real scalar can be divined from
those discussed above by realizing that a complex scalar can be
expressed as $\varphi_c= (\varphi_1+i \,\varphi_2)/\sqrt{2}$. In other words,
the relic density of the complex field may be expressed in terms of
those for the real fields as $\Omega=\Omega_{1}+\Omega_{2}$. For
identically parametrized effective Lagrangians, the vertices for the
real scalar theory would have an extra factor of 2 (as compared to
those for the complex field). On the other hand, one must account for
identical fields in the final state when a pair of DM particles is
being produced as a result of SM-pair annihilation. The consequent
change in $T_f$ is, however, only a minor one (at the level of 10\% or
lower) and has little bearing on the relic abundance calculation.
Now, $\Omega \propto \langle\sigma \, v\rangle^{-1} \sim \Lambda^n$,
where $n=2\, (4)$ for dimension-5 (dimension-6) operators.  The
aforementioned factor of 2 in the vertex, along with a factor of half
in the thermal average (owing to identical particles in the initial
state), thus, implies that the constraint on $\Lambda_{{\rm real}
  \varphi}$ would be, approximately, a factor of $2^{3/2} \, (2^{3/4})$
stronger than those derived above for dimension-5 (dimension-6)
operators. This is borne out by explicit calculations.

\subsection{Caveat to the calculation}
Until now, we have been considering the case where a pair of DM
particles, annihilate into a pair of SM particles, treating the
latter as asymptotic states. In other words, it was assumed, naively, 
that a DM-pair annihilating to hadrons could be well-approximated by 
$\varphi\,\varphi^*\rightarrow q_i\,\bar{q_i}$ with the quarks hadronizing 
subsequently. However, when the mass of the DM particle is of the order
of a quark mass, the relative momentum between the quark-antiquark pair
is small and a bound state ensues. This, obviously, would need a different
calculational scheme.  For the mass range that we are considering, this is
of relevance only in regards to the three light quarks. In particular,
note that such a DM species with interaction strengths that we are investigating
would freeze out only around the QCD phase transition temperature
(see Fig.\ref{fig:xf}). This brings with the added complication
that, even if the DM is considerably heavier than a light quark, the
annihilation products would hadronize immediately (on the scale of
the annihilation time), perhaps into a pair of bound states, thereby
utterly disallowing the approximation of quarks as quasi-asymptotic
states. Thus, if we want to admit unsuppressed DM-quark interactions, 
it is imperative that we consider annihilation to bound states, and we 
set up the formulation next.

\section{DM annihilation to various bound states}
For $m_\varphi >2m_\pi$ , the DM can annihilate to pions, kaons and other
light mesons, with more and more channels being available to heavier DM
particles. Nonetheless, if the coupling to $u/d$--quarks are
unsuppressed, the dominant effect arises from DM annihilation to
pions. The calculation of the relevant rates involves the
determination of the matrix elements of the operators for hadronic
states, and we begin by discussing these.

\subsection{Matrix Element and Form factors}
For arbitrary bound states ${\cal B}_{1,2}$, the matrix element for
the process $\varphi + \varphi^\dagger \rightarrow {\cal B}_1 + {\cal B}_2$,
driven by an operator $\mathcal{O}_I$ listed in
eq.(\ref{the_operators}), is given by
\begin{equation}
\mathcal{M}=\langle {\cal B}_1 \, {\cal B}_2 | \mathcal{O}_I  | \varphi \, \varphi\rangle = 
\frac{\mathcal{C}^f_{I}}{\Lambda} \,
        \langle {\cal B}_1 \, {\cal B}_2 | J_I^{\bar f f} | 0 \rangle \,
      \langle 0 | J^I_{\varphi} | \varphi \, \varphi\rangle \ .
\end{equation}
where the operator has been factorized into a product of two
currents. Since the DM, by definition, does not suffer strong
interactions, the vacuum saturation approximation is almost exact, a
result that we use in the second step above. Clearly, the valence
quark content of the hadrons ${\cal B}_{1,2}$ must be conjugate of each other
for the corresponding matrix element to be nonzero. In other words,
together, they should be a flavour-singlet pair. For simplicity, now
onwards, we will assume that the DM interacts universally with quarks
and leptons, and set $C_I^f=1, \, \forall f$ (though only for a given current
structure).  Wherever it is pertinent, we shall indicate the
difference that the relaxation of this assumption will entail.

The matrix elements can be parametrized in terms of form factors
multiplying momentum-dependent structures dictated by Lorentz
covariance, parity transformation and other symmetries, like isospin,
wherever applicable. Some of these are listed in
Table~\ref{tab:param}. Others can be parametrized analogously. Note that many
final state are missing in
  Table~\ref{tab:param}. For example, the $\pi\,K$ final state is
  precluded (at least to the lowest order in electroweak theory) by
  the assumption that the DM interactions do not admit flavour violation.
  Similarly, our assumption of identical couplings of the DM to the
  up-- and down-quarks implies isospin symmetry and, consequently,
  final states such as $\pi\eta$ are strongly
  suppressed\footnote{If the couplings ${\cal C}_s^u$ and ${\cal
        C}_s^d$ are unequal, this suppression is inoperative and the
      $\pi\eta$ channel opens up, with an amplitude proportional to
      $({\cal C}_s^u - {\cal C}_s^d)$. We, however, do not consider
      such an explicit $SU(2)$-violation any further.}. Indeed, as we
  shall see shortly, for scalar currents, the amplitude for this final
  state is proportional to the quark mass difference $m_u - m_d$. For
  vector currents, on the other hand, the process suffers an
  additional $v^4$ suppression, as is expected for a pair of scalars
  annihilating to another through a vector current.

\begin{table}[!ht]
\begin{tabular}{|cc||cc|}
\hline
\multicolumn{2}{|c||}{$J^P = 0^-$ : ${\cal B}_{1,2} = \pi, K, \eta, \cdots $}  &
\multicolumn{2}{c|}{${\cal B}_1 = \pi$, \quad${\cal B}_2 = \rho$}\\[0.5ex]
\hline
  $\langle 0 | \bar{q} {q} | {\cal B B}\rangle$  &  $F_s $ \hspace{0.5cm} &   $\langle 0 | \bar{q}\gamma^\sigma {q} | 
        {\pi \rho}\rangle$  & 
      $F_v$ \,$\varepsilon_{\mu\nu\rho\sigma}p_3^\mu p_4^\nu \epsilon^\rho $\\[1ex]
  $\langle 0 | \bar{q} \gamma^\sigma {q} | {\cal B B}\rangle$  &  $F_v \, q^\sigma$ \hspace{0.5cm} & & \\
\hline
\multicolumn{4}{|c|}{\bf ${\cal B}_{1,2} = \rho \, : \qquad \quad (q^\sigma = p_3 - p_4$ and 
$P^\sigma = p_3 + p_4)$ }\\
\hline
\multicolumn{2}{|c} {$\langle 0 | \bar{q} {q} |\rho \, \rho\rangle$} &\multicolumn{2}{c|} {$(F_s^1 \, m_\rho g^{\mu\nu} + F_s^2 p_3^\nu \, p_4^\mu)\epsilon_\mu(p_3)\epsilon_\nu(p_4)$}\\[1ex]
\multicolumn{2}{|c} {$\langle 0 | \bar{q} \gamma_5 {q} | \rho \, \rho\rangle$} & \multicolumn{2}{c|} {$F_p \, \epsilon^{\mu\nu\rho\sigma}p_{3\rho}p_{4\sigma}\epsilon_\mu(p_3)\epsilon_\nu(p_4)$} \\[1ex]
  
\multicolumn{2}{|c} {$\langle 0 | \bar{q} \gamma^\sigma {q} | \rho \rho\rangle$}
 & \multicolumn{2}{c|} 
     {$\dis\left[F_v^1 \, q^\sigma + F_v^2 \, (g^{\mu\sigma} P^\nu - g^{\nu\sigma} P^\mu) 
   + F_v^3 p_3^\nu \, p_4^\mu \frac{q^\sigma}{m_\rho^2} \right]
         \epsilon_\mu(p_3)\epsilon_\nu(p_4)$}
\\[1ex]
\multicolumn{2}{|c} {$\langle 0 | \bar{q} \gamma^\sigma \gamma_5{q} | \rho \, \rho\rangle$} &\multicolumn{2}{c|} {$F_a \, \epsilon^{\mu\nu\rho\sigma}
(q_\rho + P_\rho)\epsilon_\mu(p_3)\epsilon_\nu(p_4)$}\\ 
\hline
\end{tabular}

\caption{\em List of various matrix elements and their relation to form factors. }
\label{tab:param}
\end{table}

\subsection{Scalar Form Factors}
We begin by attempting to relate the simplest of the quark currents,
{\em viz.} the scalar $\bar q \, q$ to mesonic currents.  Naively, for
the DM masses of interest here, couplings to heavy quarks should not
play a role. However, they actually do, courtesy quantum
corrections. For example, integrating out the heavy quarks would
result in an effective operator of the form $\varphi^\dagger\varphi
G^{\mu\nu}G_{\mu\nu}$. At the one-loop order, this can be estimated by
calculating a triangle diagram with heavy quarks as propagators
yielding
\begin{equation}
\mathcal{O}_{\rm eff.}=\dis\sum_{i=c,b}\frac{\alpha_s}{12\,\pi\,\Lambda\,m_i}\varphi^\dagger\varphi \; G^{a\mu\nu}G^a_{\mu\nu} \ ,
\label{heavy_oper}
\end{equation}
where, for brevity's sake, higher powers in $m_i^{-1}$ have been neglected. We have also
explicitly omitted the top-quark contribution as it is highly suppressed at
this scale.

The operator in eqn.(\ref{heavy_oper}), though, suffers higher order corrections, 
and an accurate perturbative calculation thereof is rather cumbersome. It is useful, however,
to recast it in terms of the trace anomaly and appeal to the known
renormalization group flow of the energy momentum tensor
$\theta^{\mu\nu}$~\cite{Voloshin:1985tc,Chivukula:1989ze}.  In the
present context, the trace of the QCD $\theta^{\mu\nu}$ is given by
\begin{equation}
\theta_{\mu}^{\mu}= -\frac{9\alpha_s}{8 \pi}G^{a\mu\nu}G^a_{\mu\nu}  
    + \sum_{\text{light}} m_q \bar{q}q,
\end{equation}
and the operator in eqn.(\ref{heavy_oper}) becomes
\[
\mathcal{O}_{\rm eff.} = \dis\sum_{i=c,b}\frac{2}{27\,\Lambda\,m_i}\varphi^\dagger\varphi 
     \left[- \theta_{\mu}^{\mu} + \sum_{\text{light}=u,d,s} m_q \bar{q}q \right] \ .
\]
Finally, for the scalar operator, we have
\begin{equation}
\mathcal{O}_{s} = \dis\frac{\varphi^\dagger\varphi }{\Lambda} \, 
    \left(\sum_{i=c,b}\frac{2}{27\,m_i} 
    \left[- \theta_{\mu}^{\mu} + 
                               \sum_{q = u,d,s} m_q \bar{q}q\right]
       + \sum_{q=u,d,s} \bar{q}q\right) \ .
\end{equation}
Note that, if the coefficients ${\cal C}_s^q$ were different from
unity, the terms in the equation above would be trivially modified.

Using the result above, we can define the scalar form factor as
\begin{equation}
	F_s = -\sum_{k=c,b}\frac{2}{27\,m_k}\theta^\mu_\mu 
         +\left(\frac{2}{27\,m_c} + \frac{2}{27\,m_b}\right)\Gamma_m  + \Gamma
+ \left(\frac{2}{27\,m_c} + \frac{2}{27\,m_b}\right)\Delta_m + \Delta
\label{scalar_form}
\end{equation}
where
\begin{equation}
\barr{rcl}
\Gamma_m &= & \dis \langle {\cal B \, B} | m_u\bar{u}u
                                         +m_d \bar{d}d | 0 \rangle 
\\[1ex]
\Gamma &= & \dis \langle {\cal B \, B} | \bar{u}u + \bar{d}d | 0 \rangle 
\\[1ex]
\Delta_m &= & \dis \langle {\cal B \, B} | m_s \bar ss | 0 \rangle 
\\[1ex]
\Delta &= & \dis \langle {\cal B \, B} | \bar ss | 0 \rangle 
\\[1ex]
\theta &= & \dis \langle {\cal B \, B} | \theta_{\mu}^{\mu} | 0 \rangle \ .
\earr
     \label{form_facs}
\end{equation}

\subsection{Vector Form Factors}
\label{vector_ff}
For a real ({\em i.e.}, one that carries no charge or any other
additive quantum number) (pseudo)-scalar meson ${\cal B}$, a matrix element of the form $\langle 0
| \bar q \gamma_\mu q | {\cal BB}\rangle$ would, necessarily, vanish
identically. This would be the case for the $\pi^0, \eta$, but not
necessarily for the $K^0$. For the latter and for other charged
mesons, the scalar and pseudo-scalar form factor can be related to
vector and axial-vector form factor respectively through
\begin{equation}
\barr{rcl}
\dis q^\mu \langle 0 | \bar{q} \gamma_\mu {q} | {\cal B \, B} \rangle 
& = & \dis F_v \, q^2 \\[1ex]
\dis \langle 0 |  m_q \, \bar{q} {q} | {\cal B \, B}\rangle & = & \dis F_v \, q^2 
\\[1ex]
\dis \langle 0 | \bar{q} \gamma_\mu {q} | {\cal B \, B} \rangle & = & 
\dis  \frac{(F_s^{\text{light}} + F_s^{\text{heavy}}) q_\mu}{q^2} \ ,
\earr
\end{equation}
where $q^\mu = (p_{\bar q} - p_q)^\mu$. Using expressions analogous to
those in the preceding subsection to express contributions due to the
heavy quarks, the total vector form factor and hence, matrix element
for vector interaction is given by
\begin{equation} F_v = \frac{1}{q^2}\left(\frac{2}{9}\theta +\frac{7}{9}\Gamma_m + \frac{7}{9}\Delta\right) \ .
 \end{equation}
In other words, the vector form factor can be written in terms of the
scalar form factors, and the extraction of the latter
suffices. Therefore, in the rest of the paper, we will focus on 
ways to extract the scalar form factors.

An accurate determination of the form factors requires the expression
of the quark-current in terms of hadronic currents and several
approaches are possible. A particularly simple and elegant formalism
is afforded by chiral perturbation theory ($\chi$PT) and the use of
dispersion relations. Analogous techniques have been used in studying
the decay of a light scalar into
hadrons~\cite{Donoghue:1990xh,Truong:1989my,Chivukula:1989ze},and, in
the next section, we adapt these to our case.

\section{Using Chiral Perturbation Theory ($\chi$PT)}
We begin by recapitulating the key results, derived within chiral
perturbation theory ($\chi$PT), that are useful to our work. For an
in-depth discussion of the subject, numerous resources~
\cite{Leutwyler:1993iq,Bijnens:1994qh,Scherer:2002tk} exist.

As is well known, $\chi$PT describes the low energy dynamics of QCD. In
its simplest version, corresponding to the existence of just two
massless quarks ($u,d$), QCD admits an exact $SU(2)\otimes SU(2)$
chiral symmetry, and the corresponding $\chi$PT lagrangian is described by
\begin{equation}
\mathcal{L} = \frac{f_\pi^2} {4} \text{Tr} \partial_\mu U  \partial^\mu U^\dagger 
+ \frac{B f_\pi ^ 2} {2} \text{Tr} ( M^\dagger U + U^ \dagger M).
\label{2-gen_chipt_lagr}
\end{equation}
Here, $f_\pi$ is the pion decay constant, and the dynamical degrees of
freedom are encoded in a $2 \times 2$ matrix $U \in SU(2)$, which can
be parametrized as $U = e^ {i \pi(x)/f_\pi}$, where
\begin{equation}
\pi(x)  \equiv \frac{1}{\sqrt{2}}  \, 
\left(\begin{array}{cc}
\pi^0 & \sqrt{2}\pi^+\\
\sqrt{2}\pi^- & -\pi^0
         \end{array}\right)\,
\end{equation}
transforming in the adjoint representation of $SU(2)$ would be
identified with the physical pions.  In eqn.(\ref{2-gen_chipt_lagr}),
$B$ is an arbitrary coupling constant (whose physical significance is
yet to be ascertained) while $M$ is a constant $2 \times 2$ matrix to
be related to masses. The aforementioned Lagrangian would exhibit
$SU(2)_L \otimes SU(2)_R$ symmetry only if the matrix $M$ also
transforms appropriately, viz., under the action of the chiral
  symmetry group, $U \rightarrow V_L \, U \, V^\dagger_R$ and $M
\rightarrow V_L \, M \, V^\dagger_R$, where $V_{L,R}$ denote the
respective transformations under the two $SU(2)$s. Given that $M$, 
unlike $U$, is not a dynamical
variable, such a transformation may seem strange. However, note that,
for massive quarks, QCD lagrangian does not admit axial symmetry.
Indeed, even in the absence of $M$, the ground state of the above
lagrangian is not symmetric under axial symmetry. In other words, the
symmetry is spontaneously broken leading to three goldstone bosons
with odd parity. Similarly, for unequal quark masses, the $SU(2)_V$
symmetry is also lost. Keeping this in mind, a nondynamic $M$ can be
perceived as a perturbation that explicitly breaks $SU(2)_A$, thereby
rendering the pions to be only pseudo-Goldstone bosons, as also
breaking $SU(2)_V$ by a small amount proportional to the difference
$m_d - m_u$. Consequently, matrix elements can be expanded in powers
of the mass term, or, equivalently, as $\mathcal{O}(p^2)$ corrections.

For three flavours, the symmetry is enlarged to $SU(3)_L\otimes
SU(3)_R$, or, equivalently, to $SU(3)_V\otimes SU(3)_A$.  Now $U \in
SU(3)$ and $\pi(x)$ refers to the full pseudoscalar octet, {\em viz.},
\begin{equation}
\pi(x) \longrightarrow \left(
\begin{array}{ccc}
\dis \frac{\pi^0}{\sqrt{2}}+\frac{\eta}{\sqrt{6}} & \pi^+ & K^+\\[1.5ex]
\pi^- & \dis \frac{-\pi^0}{\sqrt{2}}+\frac{\eta}{\sqrt{6}} & K^0\\[1.5ex]
K^- & \dis \overline{K^0} & \dis \frac{-2 \, \eta}{\sqrt{6}}
         \end{array}
         \right)\ .
\end{equation}
Similarly, $M$, too, gets promoted to a $3 \times 3$ matrix. This is
the theory that we shall work with.

We now consider the quark operators of interest. As has been
demonstrated in the preceding section, the vector form factors can be
expressed in terms of the scalar ones. The
operators $\bar q_i q_i$ (where $q_i$ denote the light quarks) can be
expressed as
\begin{equation}
 \bar q q= - \frac{\partial \mathcal{L}_{\rm QCD} }{\partial m_q}
         =  \frac{\partial \mathcal{H}_{\rm QCD} }{\partial m_q} \ .
  \label{eq:def}
\end{equation}
On the other hand, the term, $ \text{Tr}
( M^\dagger U + U^ \dagger M)$, in the $\chi$PT Lagrangian can be
expanded up to the second order to obtain the pion, kaon and $\eta$
mass terms, namely 
\begin{equation}
\barr{rcl}
\dis \frac{f_\pi ^ 2} {2} \text{Tr} ( M^\dagger U + U^ \dagger M) 
     & = &\dis (m_u+m_d)\pi^+\pi^-+\frac{(m_u+m_d)}{2}\pi_0^2
\\[2ex]
&+ & \dis (m_u+m_s)K^+K^-+(m_d+m_s)K^0\bar{K^0} +\frac{(m_u+m_d+4m_s)}{6}\eta^2
\\[2ex]
&+ &\dis \frac{(m_u-m_d)}{\sqrt{3}}\pi^0\eta.
\earr
\end{equation}
Hence, the masses are given as
\begin{equation}
\barr{rcl c rcl}
\dis m_\pi^2 & = & \dis B (m_u + m_d), & \qquad & 
\dis m_{K^+}^2 & = & \dis B (m_u + m_s)
\\[1ex]
\dis \quad m_{K^0}^2 & = & \dis B (m_d + m_s) & & 
 m_{\eta}^2 & = & \dis B \frac{(m_u+m_d + 4m_s)}{3}.
\earr
\label{eq:meson}
\end{equation}
Using the fact that the expectation values of the respective
Hamiltonians for the two theory should be equal, we have
\begin{equation}
 \dis \Big\langle \pi \pi \Big|\sum_{q=u,d}\bar{q} q\Big|0 \Big\rangle 
				    =  B.
\end{equation}
On the other hand,
\begin{equation}
		m_\pi^2 = \hat{m} \, B \ ,  \qquad 
    \hat{m} \equiv m_u + m_d \ ,
\end{equation}
and, therefore,
\begin{equation}
 \dis \Big\langle \pi \pi \Big|\sum_{q=u,d}\bar{q} q\Big|0 \Big\rangle 
= \frac{m_\pi^2}{\hat{m}} \ .
\end{equation}
In a similar vein, we have (neglecting the difference $m_d - m_u$)
\begin{equation}
\barr{rcl c rcl}
 \dis \Big\langle K^+ K^- \Big|\sum_{q=u,d}\bar q q\Big|0 \Big\rangle 
& = & \dis \frac{m_{\pi}^2}{2\, \hat{m}} \ , 
&\qquad & 
\dis \langle K^+ K^- |\bar s s|0 \rangle & = & 
   \dis \frac{2 m_{K^+}^2-m_{\pi}^2}{2 \, m_s} \ ,
\\[1.5ex]
\dis \Big\langle \eta\eta \Big|\sum_{q=u,d}\bar q q\Big|0 \Big\rangle 
&= & \dis \frac{m_{\pi}^2}{3\, \hat{m}} \ , 
& & 
 \dis \langle \eta\eta |\bar s s|0 \rangle &= & \dis
\frac{3\,m_{\eta}^2 - m_{\pi}^2}{4 \, m_s} \ ,
 \earr
 \end{equation}
where the matrix elements for the $\bar s s$ current are obtained 
by differentiating ${\cal L}_{\rm QCD}$ with respect to $m_s$.

At this order, the trace of the energy momentum tensor reads
\begin{equation}
\theta_\mu^\mu = - \partial_\mu \pi \partial^\mu \pi + 2 m_\pi^2 \pi\cdot\pi
\end{equation}
Therefore, the corresponding form factors (see eqn.(\ref{form_facs})
for definition) are rendered
\begin{equation}
\barr{rcl c rcl c rcl c rcl}
\theta_\pi(s) &=& s+2m _\pi^2 \ , & \quad &
\Gamma^\pi_m(s) &= & m_{\pi}^{2} \ , & \quad & 
\Gamma_\pi(s) & = & \dis \frac{m_{\pi}^{2}}{\hat{m}}\ ,
& \quad &  \Delta_\pi(s) & = & 0 \\[1.5ex]
\theta_K(s)& =& s+2m _K^2 \ , & & 
\Gamma^K_m(s) & = & \dis \frac{ m_{\pi}^{2}}{2} \ , & & 
\Gamma_K(s) & = & \dis \frac{m_{\pi}^{2}}{2\hat{m}} \ , & & 
 \Delta_K(s) & = & \dis \frac{2\, m_{K^+}^2 - m_{\pi}^2}{2 \, m_s}\\[1.5ex]
\theta_\eta(s) &= &s+2m _\eta^2 \ , & & 
\Gamma^\eta_m(s) & = & \dis \frac{ m_{\pi}^{2}}{3} \ , & & 
\Gamma_\eta(s) & = & \dis \frac{m_{\pi}^{2}}{3\hat{m}}, & & 
\Delta_\eta(s) & = & \dis 
\frac{3 m_{\eta}^2 -  m_{\pi}^2}{4 \, m_s}
\earr
       \label{formfac_chPT_0}
\end{equation}
and, consequently, the total matrix element becomes
\begin{equation}
	F_{s} =\frac{1}{\Lambda}\left[-\sum_{k=c,b}\frac{2}{27\,m_k}\theta 
+\left(\frac{2}{27\,m_c} + \frac{2}{27\,m_b} 
+ \frac{1}{\hat{m}}\right)\Gamma_m + \Delta_s\right] \ .
\end{equation}
It should be noted that, in the limit of isospin symmetry ($m_u=m_d=\hat{m}/2$), 
the various scalar form factors obey
\begin{equation}
\barr{rclcl}
\Gamma_{\pi,s}(s) &\equiv&  
        2 \Gamma^{\pi^0\pi^0}_{uu}(s) &=&
        2 \Gamma^{\pi^0\pi^0}_{dd}(s) =
        2 \Gamma^{\pi^+\pi^-}_{uu}(s) =
        2 \Gamma^{\pi^+\pi^-}_{dd}(s) 
\\[1ex]
\Delta_{\pi}(s) &\equiv & \Delta^{\pi^+\pi^-}_{ss} &= &\Delta^{\pi^0\pi^0}_{ss}\,.
\earr
\end{equation}
Similar expressions hold for the kaonic form factors. Using these form
factors and the expressions for the cross sections, we may obtain the
relic abundance. 

It should be realized, though, that the form factors derived so far
have been defined within the lowest order $\chi$PT. This is reflected by
the form factors being constants rather than functions of momenta. In
other words, the aforementioned values only reflect the values of the
form-factors at a particular momentum scale, defined by the
decay/interaction which these are extracted from. As we shall shortly
see, the higher-order corrections can be quite
important. Consequently, we postpone the calculation of the relic
abundance until after at least some of these corrections are
evaluated.

The $\chi$PT Lagrangian can be expressed as a power series in the exchange
momentum $p$; terms containing quark masses or external scalar or
pseudo scalar fields are ${\cal O}(p^2)$ whereas external vector or
axial-vector fields are ${\cal O}(p)$.  The NLO terms in the
Lagrangian contains terms that are ${\cal O}(p^4)$ or, in other words,
suppressed by further factors of ${\cal O}(p^2/\Lambda_{\rm QCD}^2)$.
With $\Lambda_{\rm QCD}\sim200\mev$, clearly, a perturbative
calculation of the higher-order effects is valid only for small
momentum exchanges. In the present context, this translates to a limit
on the dark matter mass, viz. $m_\varphi\lapp 300
\mev$~\cite{Chivukula:1989ze}, for a perturbative expansion to make
sense.  Instead, we calculate the form factors using dispersion
relations as this method relies solely on general principles and data.

\subsection{Form factors from Dispersion Relations}

We now discuss how the form factors can be extracted from scattering
data using dispersion relations. This would allow us to determine the
deviations, as compared to the preceding section, wrought in the relic
abundances. Since the pion and kaon final states result in the
dominant contributions, we would be concentrating primarily on these
two form factors. In this, we largely follow the methodology developed
in Refs.\cite{Donoghue:1990xh,Ananthanarayan:2004xy}. 

\subsubsection{From $\chi$PT to Dispersion Relations} 
\begin{figure}[!h]
\centering
\includegraphics{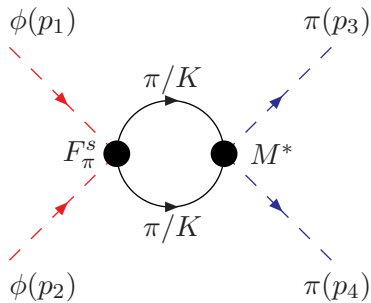} 
\caption{\em Typical diagrams that contribute to $\varphi\varphi \to \pi\pi$ at 
the next to leading order.}
\label{fig:cutkosky}
\end{figure}

With interactions amongst the hadrons switched on, at the one-loop
level, diagrams as in Fig.\ref{fig:cutkosky} would also
contribute. While we have denoted only a subset of the one-loop
diagrams, multiple intermediate states do contribute. And, with the
hadron-hadron interactions being strong, there is no {\em a priori}
compelling reason to limit ourselves to only one-loop results. In
other words, to write down the S-matrix for such a system, we need to
include contributions from channels like $\pi\pi\rightarrow\pi\pi$,
$\pi\pi\rightarrow K^+\,K^-$, $\pi\pi\rightarrow 4\pi$,
$\pi\pi\rightarrow \eta\,\eta$ etc.  The direct calculation of the
loops is, of course, a very difficult task. Instead, we take recourse
to determining the imaginary part (wherever applicable) using the
Cutkosky rules and, subsequently, calculating the real part using
dispersion relations. Although, all channels do contribute to the
total amplitude, in this work, we are restricting to two channels
alone, viz.  $\pi\pi$ and $K\bar{K}$ as these are expected to
overwhelmingly dominate upto
$m_{\varphi\varphi}=1.4\gev$\cite{Moussallam:1999aq} {\em i.e.,} for dark
matter of mass $\leq$ 700\mev. Above $m_{\varphi\varphi}=1.4\gev$, states
such as $4\pi$ and $\eta\eta$ come into play, with the contribution of
$f_0(1500)$, in particular, expected to be felt beyond $m_{\varphi\varphi}=1.5\gev$
\cite{Moussallam:1999aq}. Nonetheless, the 2-channel approximation is
expected to be a very good one, as the other channels ($X$) mentioned
above typically suffer from either kinematic restrictions (leading to
a vanishing imaginary part) or small $\varphi\varphi \to X$
amplitudes\footnote{In addition, for certain choices of $X$, even the
  amplitude for $\pi\pi$ to $X$ is suppressed as well.}. We shall see
this explicitly in results below.

Given that the initial state ($\varphi\varphi$) is well-described by $J^P =
  0^+$, we are interested only in final states with $I = 0$ and $J^P =
  0^+$. Under the assumption that the other channels can be entirely
  neglected, the  S-matrix for meson scatterings 
  (which lives in the aforementioned subspace) 
  can be further reduced to a $2\times2$ unitary
  submatrix given by 
\begin{equation}
 S_{jk} = \delta_{jk} + 2i\sqrt{\sigma_j\sigma_k}M_{jk} \ , \qquad 
     \sigma_j \equiv \sqrt{1-4\,m_j^2/s}
 \label{eqn:s1}
\end{equation}
where $j,k=\pi\, , K$ and $M_{jk}$ is the corresponding element of the
transition matrix.

 Using the unitarity of the S-matrix, the imaginary part of the
 transition matrix can be expressed as
\begin{equation}
 \text{Im}(M_{ij}) = \sum_{k} M^*_{ik}\,M_{kj}\sigma_k\Theta(s-4m_k^2),  
\end{equation}
and, similarly, for the imaginary part of form factor viz.,
\begin{equation}
 \text{Im}(F^i_s) = \sum_{k} M^*_{ik}\,F^k_s\sigma_k\Theta(s-4m_k^2), 
 \label{eqn:s2}
\end{equation}
with $\Theta(x)$ being the well-known step function.

More explicitly, the Cutkosky rules determine the discontinuity, and,
hence, the imaginary part of the scattering amplitude through the
Schwarz reflection principle. In doing this, it needs to be realized
that\cite{Donoghue:1990xh} the form factors of eqn.(\ref{form_facs}),
treated as functions of $s \equiv p^2$, where $p_\mu$ is the momentum
transfer, are analytic in the complex $s$-plane, except for a cut
along the positive real axis. For the two-meson (${\cal BB}$) case,
the cut starts at $s = 4 m_{\cal B}^2$. While for $s > 4 m_{\cal
  B}^2$, the identifications in eqn.(\ref{form_facs}) hold, for $0 < s
< 4 m_{\cal B}^2$, the function $\theta_{\cal B}(s)$ represents the
matrix element $\langle 0 |\theta_\mu^\mu|{\cal
  BB}\rangle$. Similarly, for real and negative values of $s$, it
corresponds to $\langle {\cal B} |\theta_\mu^\mu|{\cal B}\rangle$ and
is real. This, in turn, implies that the values above and below the cut
are complex conjugates of one another.

Consider, for example, the region $4 m_\varphi^2 < s < 4 m_K^2$, the
only allowed process for DM annihilation is, thus\footnote{Here, we
  include the possibility that $\varphi$ may represent a real
  scalar. Similarly, $\pi\pi$ includes both $\pi^+\pi^-$ and $\pi^0
  \pi^0$.} $\varphi \varphi^* \to \pi \pi$, with $3\pi$ final states being
precluded by isospin invariance. Consequently, the only allowed final
state rescattering is $\pi\pi \to \pi\pi$, and, in the limit of
identical masses for the charged and neutral pions, is an entirely
elastic process. Concentrating on the rescattering, the in- and
out-states only differ in phase. For such a single-channel case, if we
denote the form-factor on the upper side of the cut by $F^\pi_s$, then
  \begin{equation}
      F^\pi_s = S_{\pi\pi} F^{*\pi}_s     \ ,
     \qquad S_{\pi\pi} = \exp\left[2 \, \pi \, i \, \delta_\pi(s) \right],
\end{equation}
where $\delta_\pi(s)$ is the $I=0, \, J = 0$ pion-scattering phase
shift. Consequently, as the cut is approached from the above, $F^\pi_s
\, \exp\left(i \, \pi \, \delta_\pi \right)$ is a real quantity.

Once $\delta_\pi$ is known (from data), what remains is to determine 
$F^\pi_s$.  If, for $s\rightarrow\infty$, the phase shift $\delta_\pi$ 
tends to a finite value and $F^\pi_s$ does not grow faster than 
a power of $s$, then the form factor is known to be given 
by the Omn\`{e}s function\cite{Omnes:1958hv} $\Omega(s)$ as
\begin{equation}
 F_s(s) = P(s)\Omega(s) = P(s) \, \exp\left(\frac{s}{\pi}\int^\infty_{4m_\pi^2} \,
                \frac{ds'}{s'}\frac{\delta_\pi(s')}{s'-s-i\epsilon}\right)
\end{equation}
where $P(s)$ is a polynomial to be fixed by the behaviour of $F_s(s)$.
It is straightforward to prove that, for
$\delta(s\rightarrow\infty)\rightarrow \alpha \pi$, the Omn\`{e}s
function is monomially suppressed, namely,
$\Omega(s\rightarrow\infty)\rightarrow s^{-\alpha}$. Moreover, the
high energy behaviour of QCD dictates that, asymptotically, the form
factor behaves as $s^{-1}$. So, for pion-pion scattering, where phase
shift at $s=\infty$ goes as $\pi$, the Omn\`{e}s function would go as
$1/s$ and $P(s)$, immediately, is constrained to be a constant, to be
determined from the value of 
form factor at $s=0$.

For larger values of $s$, further channels come into play.
Restricting ourselves, as argued for above, to
 two channels, the expression above is generalised to
  \begin{equation}
F^i_s = \sum_{j=0}^1 (\delta_{ij} + 2i\sigma_jM_{ij}) F^{*j}_s  
\label{eqn:s3}
  \end{equation}
where the interaction amplitude for a process can be expanded in
partial waves with angular momentum $l$:
\[
M = \dis\frac{1}{2i\sigma}\sum_{l=0}^l (2l+1)(e^{2i\delta_l} -1)P_l(\cos\theta)
\]
For scalar form factors, we need to consider only the $l=0$ part.
Taking a cue from the single-channel case, the S-matrix for two
channel process can be parametrized as
\begin{equation}
S_{\rm total} =\left(
\begin{array}{cc}
\cos\theta \, e^{2i\delta_\pi} & i \, \sin\theta \, e^{i(\delta_\pi+\delta_K)}\\
i \, \sin\theta \, e^{i(\delta_\pi+\delta_K)} & \cos\theta \, e^{2i\delta_K}
\end{array}
         \right)\,
\end{equation}
where $\cos\theta$ determines the mixing between the
two channels.  Clearly, for $\cos\theta=1$,
the two channels decouple entirely, and the solutions are to be 
obtained independently from $\pi\pi \to \pi\pi$ and $KK \to KK$ respectively,
namely
\[ F^\pi(s)= F^1(s) = P_1(s) \Omega^1_1(s)  \qquad F^K(s)=F^2(s) = P_2(s) \Omega^2_2(s) \ .
\]
However, as is expected, and as can be ascertained from data (see, for
example, Fig.4 of Ref.\cite{Au:1986vs}), $\cos\theta \neq 1$. 
Non-trivial values of $\cos\theta$ essentially parametrize the
relative strength of, say, $KK$ admixture in the determination of the
$\pi\pi \to \pi\pi$ scattering. Known as the elasticity
parameter, $\cos\theta$ is a function of energy, angular momentum and
isospin. Along with the phase shifts, we treat $\cos\theta$ as an
experimental input.

It is useful to parametrize the consequent form factors through
\begin{equation}\label{eq:omnessolution}
     \begin{pmatrix}
 F^\pi(s) \\[1ex]   
 \frac{2}{\sqrt{3}}F^K(s)
    \end{pmatrix}= \begin{pmatrix}
 \Omega^1_1 &&\Omega^1_2 \\[1ex]    
 \Omega^2_1 &&\Omega^2_2
    \end{pmatrix}
    \begin{pmatrix}
 F^\pi(0) \\[1ex]    
 \frac{2}{\sqrt{3}}F^K(0)
    \end{pmatrix}
\end{equation}
where the Clebsch-Gordan coefficient occurring in the projection of
the $\pi\pi$ state with $I=0$ is shifted to $F^K$, and can be thought
of as a relative normalisation.

\subsubsection{The two channel solution: iterative procedure} 

\begin{figure}[htb]
\centering
 \subfigure[]{\label{fig:formfac_disp_pi}\includegraphics[width=75mm]{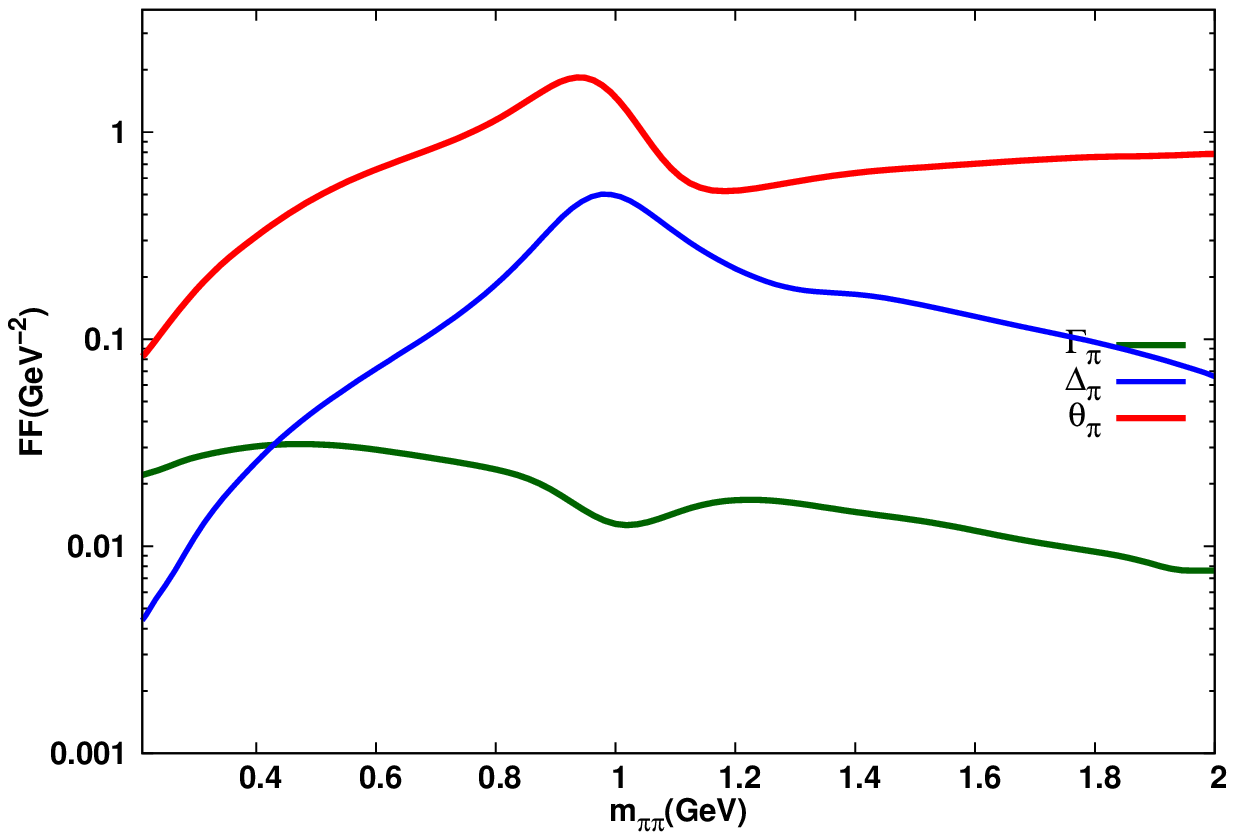}}
 \subfigure[]{\label{fig:formfac_disp_k}\includegraphics[width=75mm]{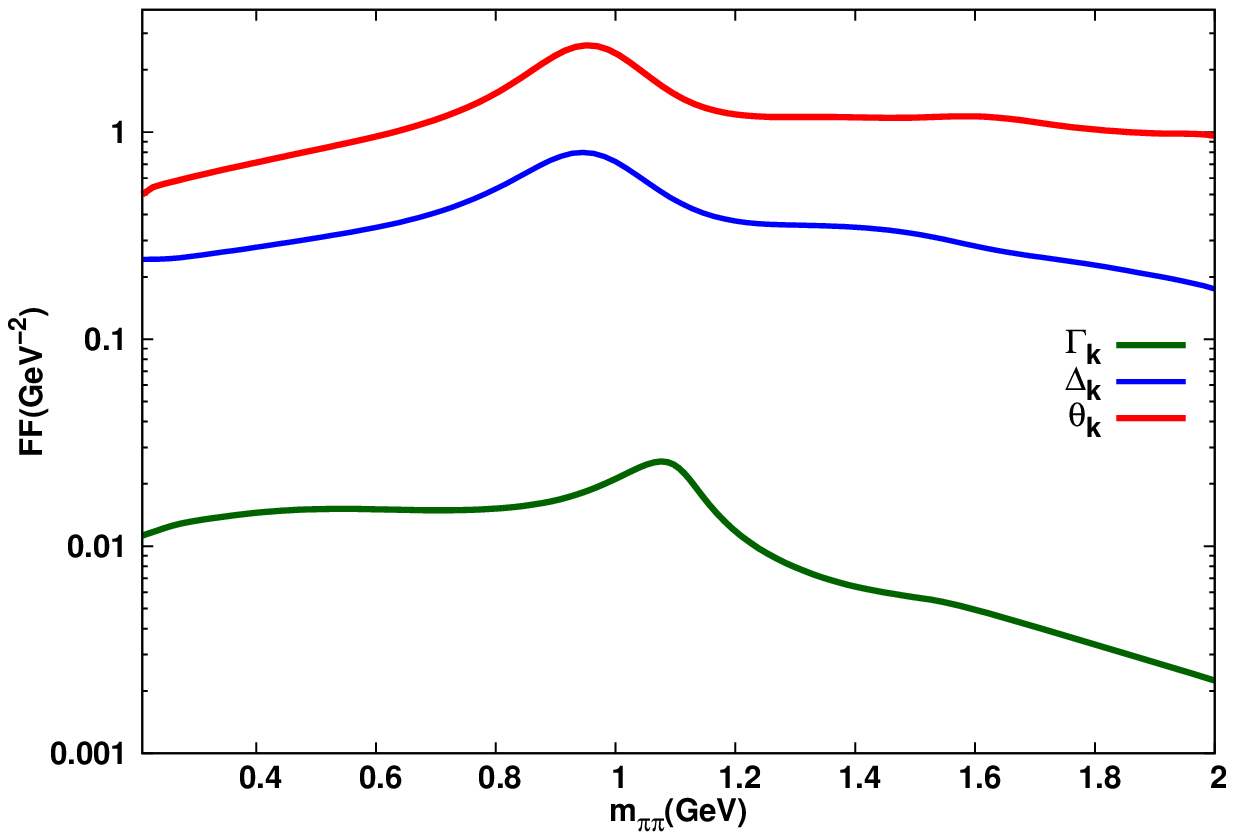}}
  \caption{\em The $s \, (=4 m_{\pi\pi}^2)$-dependence of various form factors associated with 
    {\em (a)} the pion and {\em (b)} the kaon.}
  \label{fig:formfac_dispersion}
\end{figure}

The parameters of the S-matrix (or, equivalently, the T-matrix) may be
determined using various scattering data and certain theoretical
constraints in the Roy-Steiner equations\cite{Roy:1971tc}. (An easier
route would be use the existing determination of phase shifts and
inelasticity parameters, such as those found in
Refs.\cite{Hoferichter:2012wf, Au:1986vs}.  Before we describe the
calculation of $\Omega^i_j$'s, let us express the form factors in
terms of their values at low momentum transfers $(s = 0)$, viz.
\begin{equation}
\barr{rcl} \Gamma_\pi(0) &=& \dis m_\pi^2\left(\Omega^1_1 + \frac{1}{\sqrt{3}} \Omega^1_2\right)\\[2ex]
 \Delta_\pi(0) & =& \dis  \frac{2}{\sqrt{3}}\left(m_K^2-\frac{m_\pi^2}{2}\right)\Omega^1_2\\[2ex]
 \theta_\pi(0) & =& \dis  \left(2 m_\pi^2 + p_1 s\right)\Omega^1_1+  \frac{2}{\sqrt{3}}\left(2m_K^2 + p_2 s\right)\Omega^1_1\\[2ex] 
  \Gamma_K(0) & =& \dis  \frac{m_\pi^2}{2}\left(\sqrt{3}\,\Omega^2_1 + \Omega^2_2\right)\\[2ex]
 \Delta_K(0) & =& \dis  \left(m_K^2-\frac{m_\pi^2}{2}\right)\Omega^2_2\\[2ex]
  \theta_K(0) & =& \dis  \frac{\sqrt{3}}{2}\left(2 m_\pi^2 + p_1 s\right)\Omega^2_1 +  \left(2m_K^2 + p_2 s\right)\Omega^2_2.
\earr
\end{equation}
Similarly, $\Omega^i_j$ are normalised as $\Omega^1_1(0) =
\Omega^2_2(0) =1$ and $\Omega^1_2(0) = \Omega^2_1(0) =0$. The
parameters $p_i$ are related to the slopes ($d \theta_i / d s$ and $d
\Omega^i_j / d s$) evaluated at $s=0$.  Relatable to $SU(3)$ breaking
effects, and estimated by the use of unsubtracted dispersion
relations, these lie in the range $(0.9,1.1)$. In other words, $p_i =
1$ represents a very good approximation and reproduces the
zeroth-order relations of eqn.(\ref{formfac_chPT_0}).  Armed with
these ``initial conditions'', we follow an iterative procedure similar
to that outlined in Refs.\cite{Donoghue:1990xh,Ananthanarayan:2004xy},
to calculate the form factors.  It is convenient to begin with, say,
$\Delta_\pi$ and $\Delta_K$, as these can be handled in a fashion
similar to the one channel case.  We confirmed that, asymptotically
these two solutions indeed go as $s^{-1}$. Furthermore, in the zeroth
approximation, it is assumed that form factors behave as $F^\pi(s)
=1$, $F^K(s) =\lambda$ where $\lambda$ is a real number. Then, the
imaginary part of $F^i(s)$ at every iteration is computed via eqn. 
\ref{eqn:s2} and the real part is computed using
\[ \text{Re}[F^{i(n)}(s_o)] = \dis \frac{1}{\pi}\int^\infty_{4m_\pi^2}ds\,\frac{\text{Im}F^{i(n-1)}(s)}{s-s_o}
\]
The nested integral equations can be evaluated using standard
numerical methods (see, {\em e.g.}, Ref.\cite{Moussallam:1999aq}).
 The resultant form factors, as displayed in Fig.~\ref{fig:formfac_disp_pi}
 and Fig.~\ref{fig:formfac_disp_k}, are consistent with those obtained
 in Ref.\cite{Monin:2018lee,Winkler:2018qyg}. 

\subsection{Revisiting the Relic abundance for scalar interaction}
\begin{figure}[htb]
\centering
 \subfigure[]{\label{b1}\includegraphics[width=75mm]{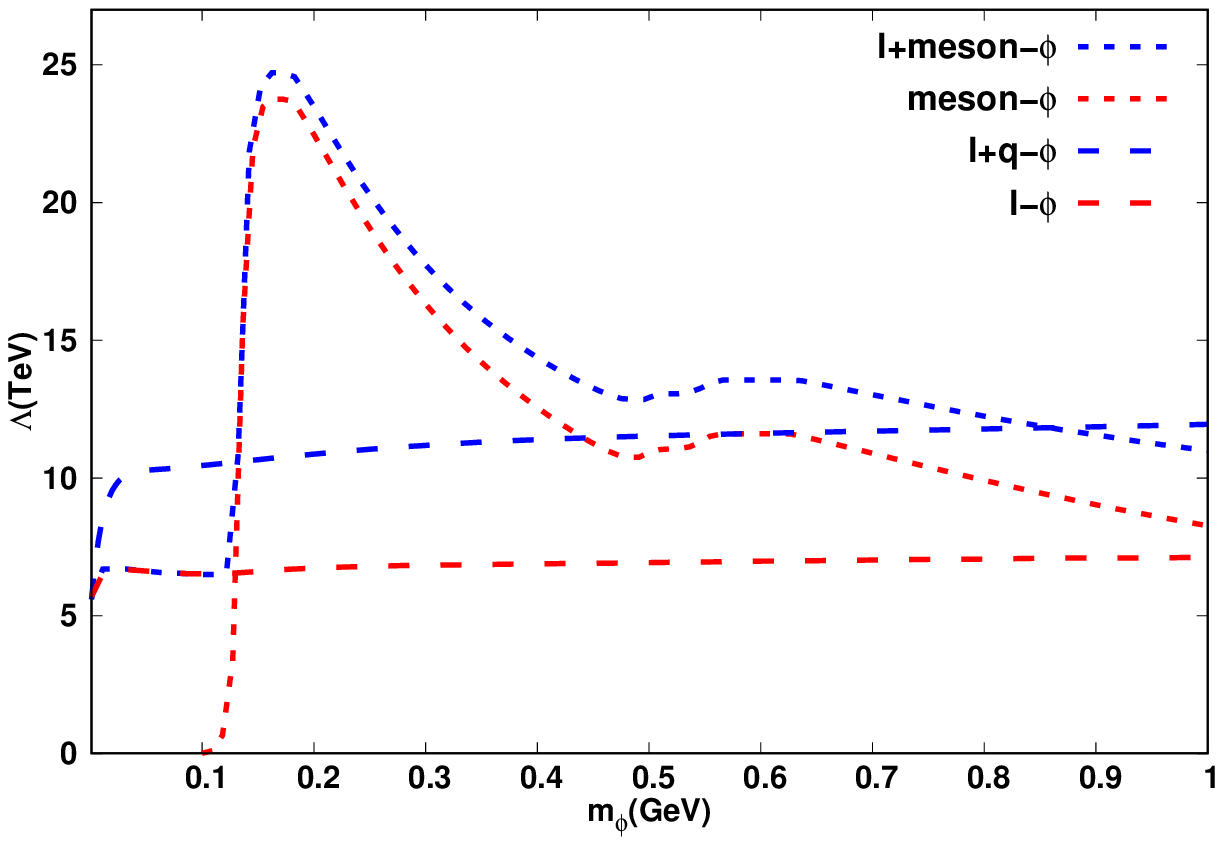}}
  \subfigure[]{\label{b2}\includegraphics[width=75mm]{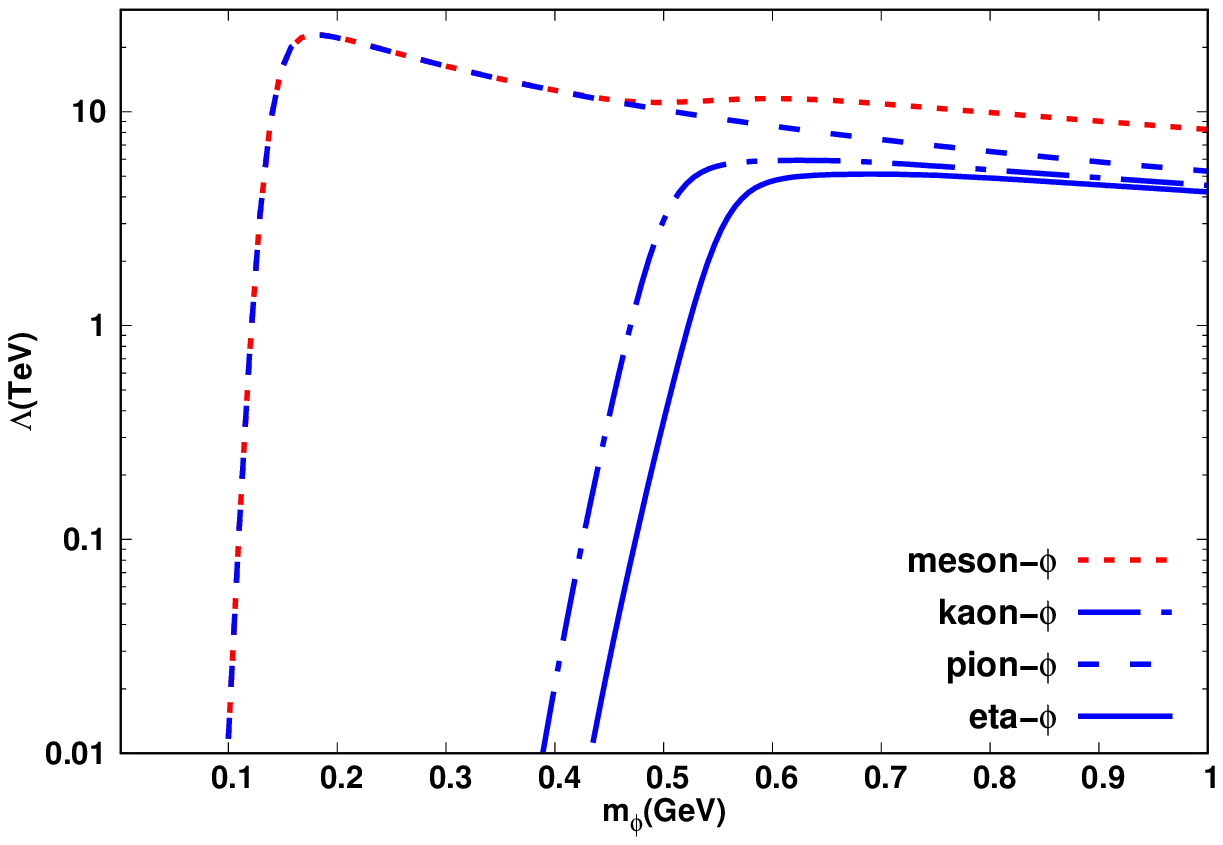}}
  \caption{\em {\em (a)} Contours in the $m_{\varphi}-\Lambda$ plane 
    (for the ${\cal O}_s$ operator in eqn. \ref{the_operators})
    with form factors determined using 
	  the lowest order chiral perturbation theory results;
          illustration of how the required $\Lambda$ changes as more 
            pseudoscalar states are included in the analysis.}
\label{fig:chiralPT} 
\end{figure}

Having obtained the form factors for scalar interaction, we have, once
again, plotted the contours in the $m_{\varphi}-\Lambda$ plane in
Fig.\ref{fig:chiralPT} using form factors determined at the
lowest order in $\chi$PT. Whereas Fig.\ref{fig:relicdensity} corresponds
to the evidently untenable assumption that the quarks may be treated as free
particles, Fig.\ref{fig:chiralPT} reflects the inclusion of bound state
effects. One particular effect was expected. Contrary to the previous case,
the annihilation channel into quarks open up only for $m_\varphi \sim m_\pi$. 
This was only to be expected as the pion is lightest of the hadrons. 
That values of $m_\varphi$ slightly smaller than $m_\pi$ are allowed too, is ascribed to the fact that 
the DM does have a non-zero momentum, albeit, somewhat smaller than its mass.

More interesting is the height of the peak. This sudden change is
to be understood in terms of the structure of the annihilation cross
sections for the different channels. For annihilations into a pair of
  leptons, the rate is proportional to $\Lambda^{-2}$. While this scaling
  originates from the structure of the effective Lagrangian in 
  eqn.(\ref{the_operators}) and remains operative for annihilations
  into mesons as well, now one has to include the effect of the nontrivial
  wavefunction of the bound state. For a two-pion final-state, this appears in the matrix
  element as $F_s(q^2)$ which has mass dimension 2, and the corresponding cross section would
  scale as $\sigma \propto m_\pi^4/(\Lambda^2 \hat{m}^2 \hat s)$. With
  $\hat s \sim 4 m_\pi^2$. it is the smallness of $\hat m$ that pushes
  up the cross section near the threshold. A similar bump exists for $m_\varphi$ slightly larger than $m_K$,
  but with a much smaller amplitude owing to the analogue of $\hat m$
  being dominated by the much larger $m_s$.  Even smaller is the
  contribution from the $\eta$-meson (as is testified to by
  Fig.\ref{b2}). However, note that, for $m_\varphi \sim 1 \gev$ the
  curve for the lowest-order bound-state analysis (wherein we have
  used only the $\pi\pi, KK, \eta\eta$ final states) is already
  veering very close to the free-quark one (which has been provided in
  Fig.\ref{b1} for reference). Clearly the inclusion of more bound states will, 
  asymptotically, render the contours very close to each other. For $m_\varphi \lapp
  1 \gev$, on the other hand, the true annihilation cross-sections
  are, typically, larger than what what transpires for free quarks,
  and, hence, the preferred value of $\Lambda$ somewhat larger.

\begin{figure}[htb]
\centering
 \subfigure[]{\label{f0}\includegraphics[width=75mm]{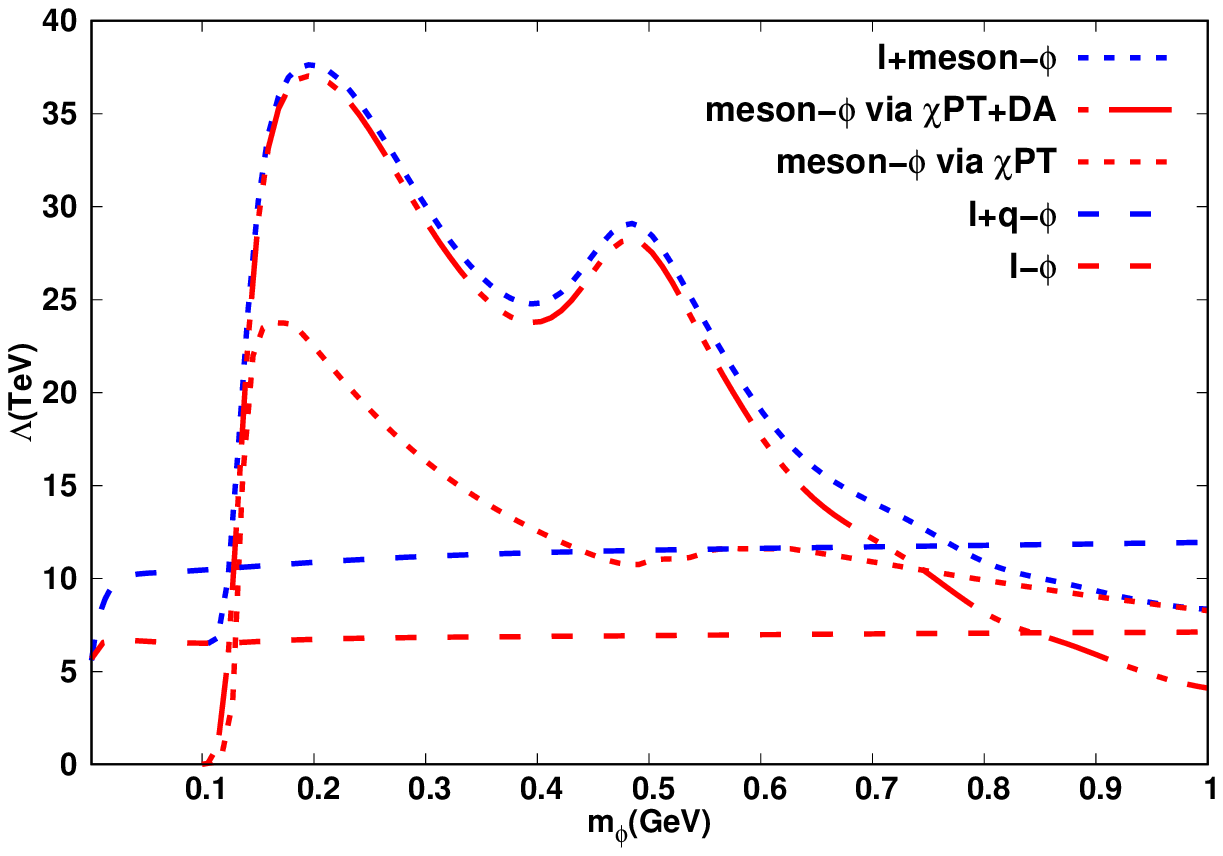}}
   \subfigure[]{\label{f1}\includegraphics[width=75mm]{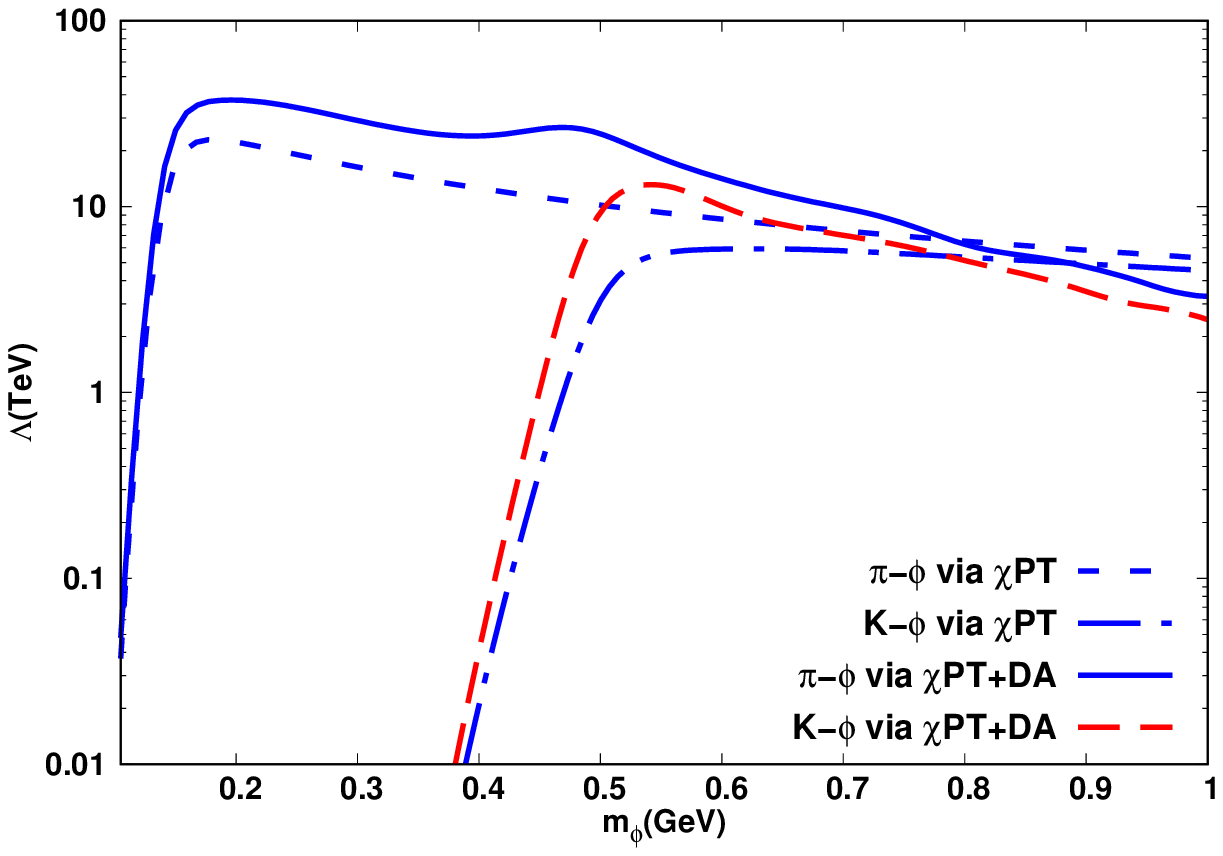}}
  \caption{\em Contours in the $m_\varphi$-$\Lambda$ plane for dimension-5 operators satisfying $\Omega_{\varphi}h^2 =$ 0.1199 $\pm$ 0.0022, 
  obtained using form factors determined by Chiral Perturbation theory and dispersion analysis(DA)}
  \label{fig:dispersion}
\end{figure}

  We now move on, from the lowest order analysis, to the inclusion
  of higher orders, through the use of dispersion
  relations. Fig.\ref{fig:dispersion} displays the corresponding
  contours. As is evident from a comparison with Fig.\ref{fig:chiralPT},
  the gross behaviour is quite similar. Overall,
  the increase in the sizes of the form factors, that the higher-order
  terms entail, results in a further rise in the preferred value of
  $\Lambda$. The bump around $m_\varphi \approx m_K$ is to be understood
  in terms of the interference between the pion and kaon form factors whose details
are secured in the data of pion and kaon phase shifts.
\normalsize

\subsection{Vector Interactions}
For a final state comprised of a pair of (pseudo)scalars, 
the form factors for a vector current can be calculated 
in terms of those for the corresponding scalar current.
As discussed in Sec.\ref{vector_ff}, such annihilation cross sections
suffer a $v^4$ suppression and, as a result, for vector
interactions, such final states contribute very little 
(as compared to, say, the leptons). On the other hand,
were the final state to comprise, say a pion and a rho meson
or a pair of vector mesons, the cross sections would no
longer be suppressed, and the conclusions would change
drastically. In view of this, we next make an estimate of
the time like form factors for these states. 

In doing this, we would be using results pertaining to the
well-studied electromagnetic current, which, for light mesons,
reads
\[
\sum_q \langle 0 | e_q \bar{q} \gamma_\mu {q} | {\cal B}_1 \, {\cal B}_2\rangle = 
\frac{2}{3} \, \langle 0 | \bar{u} \gamma_\mu {u} | {\cal B}_1 \, {\cal B}_2\rangle
- \, \frac{1}{3} \langle 0 | \bar{d} \gamma_\mu {d} | {\cal B}_1 \, {\cal B}_2\rangle.
\]
Assuming isospin symmetry, we further have
\[
\langle 0 | \bar{d} \gamma_\mu {d} | {\cal B}_1 \, {\cal B}_2\rangle=
 \langle 0 | \bar{u} \gamma_\mu {u} | {\cal B}_1 \, {\cal B}_2\rangle
= 3 \, \sum_q \langle 0 | e_q \bar{q} \gamma_\mu {q} | {\cal B}_1 \, {\cal B}_2\rangle
= 3 \,F_{\rm em}(q^2) \, \varepsilon_{\mu\nu\sigma\omega}p_1^\nu p_2^\sigma \epsilon^{*\omega}
\]
and, finally,
\begin{equation}
\sum_q 
\langle 0 | \bar{q} \gamma_\mu {q} | \pi(p_1) \, \rho(p_2, \lambda)\rangle=
6 \, F_{\rm em}(q^2) \, \varepsilon_{\mu\nu\sigma\omega}p_1^\nu p_2^\sigma \epsilon^{*\omega}(\lambda),
\end{equation}
where $\lambda$ denotes the polarization state of the $\rho$-meson and
$F_{\rm em}(q^2)$ is the electromagnetic form factor. The other form factors
discussed in Table \ref{tab:param} can be defined analogously. 

While, for the $\rho\rho$ final state, the results of  Ref.\cite{deMelo:2016lwr}
can be used in a relatively straightforward manner, the $\pi\rho$ state 
needs more work, especially in determining the time-like electromagnetic 
form factor in the region $q^2\lapp 4\gev^2$. Towards this end, we make use 
of the vector meson dominance model,wherein, for $q^2\lapp 4\gev^2$, 
the major contributions accrue from the $\omega(782)$, $\phi(1020)$ and 
$\omega(1420)$.

Restricting ourselves, for the time being, to a single vector meson
$\hat V_\mu$, the part of the Lagrangian governing its propagation,
and the interactions with an external current (such as that
corresponding to a pionic current) $J^\mu$ can be parametrized
by~\cite{Sakurai,OConnell:1995nse}
\begin{equation}
{\cal L} = - \, \frac{1}{4}\hat F_{\mu\nu} \hat F^{\mu\nu} 
           - \, \frac{1}{4}\hat V_{\mu\nu} \hat V^{\mu\nu}
           - \, \frac{\hat e}{2 \, \gv} \, \hat F_{\mu\nu} \hat V^{\mu\nu} 
           + \frac{1}{2} \, m_V^2 \, \hat V_\mu \hat V^\mu
           - \hat e \, \hat A_\mu J^\mu - \gjv \, \hat V_\mu \, J^\mu \ .
\label{VMD_1}
\end{equation}
Here, $\hat V_{\mu\nu}$ and $\hat F_{\mu\nu} $ are the usual
  field strengths corresponding to $\hat V_\mu$ and the photon field
  $A_\mu$. While $g_{{}_{VJ}}$ determines the strength of the
  coupling, the term containing $g_{{}_V}$ parametrizes the kinetic
  mixing (between the two vector fields) that is allowed in an abelian
  theory.  The presence of this last term (which could also be thought
  of as a momentum-dependent $\gamma$--$\rho$ coupling vanishing at
  $q^2 = 0$) calls for a re diagonalization of the kinetic part
  (including the mass term) of the Lagrangian so as to permit usual
  perturbative analysis. While this would be a standard exercise, a
  slightly different formulation is more
  common~\cite{Sakurai,OConnell:1995nse}, namely
\begin{equation}
\hat e \to e \equiv \hat e \, \sqrt{1 - \, \frac{\hat e^2}{\gv^2}}
\ , \quad
\hat A_\mu \to A_\mu \equiv \frac{e}{\hat e} \, \hat A_\mu 
\ , \quad 
\hat V_\mu \to V_\mu \equiv \hat V_\mu + \frac{\hat e}{\gv}  \, \hat A_\mu \ .
   \label{VMD_field_redefn}
\end{equation}
For the special point $\gjv = \gv$ (necessary to maintain $F_{\rm em}(q^2 =
  0) = 1$), this transformation leads to, approximately,
\begin{equation}
{\cal L} = -\, \frac{1}{4} \hat F_{\mu\nu} \hat F^{\mu\nu}
           -\, \frac{1}{4} \hat V_{\mu\nu} \hat V^{\mu\nu}
+\frac{1}{2}m_V^2 \hat V_\mu \hat V^\mu-\frac{e\, m_V^2}{gv} \hat V_\mu \hat A^\mu
+\frac{e^2 m_V^2}{2\,\gv^2}\hat A_\mu \hat A^\mu - \gjv \hat V_\mu J^\mu \ ,
   \label{VMD_2}
\end{equation}
with the difference between the two Lagrangians being ${\cal O}(e^3 /
\gv^3)$.  The existence, in eqn.(\ref{VMD_2}), of a mass term for the
photon is, of course, a concern. However, note that there now exists a
mass-mixing between the photon and the $V$, in lieu of the earlier
kinetic mixing. Summing to all orders in this mixing (no loops, though), 
the tree-level photon propagator can be easily seen to be proportional to

\[
\frac{1}{q^2} \, \left[1 + \frac{e^2 m_V^2}{\gv^2 \, (m_V^2 - q^2)} \right]^{-1} \ .
\]
While the $q^2 \to 0$ limit reproduces the standard propagator (modulo
  a renormalization), the existence of the second pole is a reminder
  of the fact that the field redefinitions in
  eqn.(\ref{VMD_field_redefn}) are neither complete nor even
  unitary. Neglecting this aspect for the time being, it is easy to
  see that, for $\gamma^* \to \pi^+ \pi^-$, while the Lagrangian of
  eqn.(\ref{VMD_1}) trivially reproduces $F_{\rm em}(q^2 = 0) = 1$,
  the one in eqn.(\ref{VMD_2}) would have done so only for $\gjv
  = \gv$, as claimed earlier.  With the absence of any direct
  coupling of the photon with the current $J^\mu$ existing in
  eqn.(\ref{VMD_2}), the latter represents, explicitly, the situation
  of complete vector meson dominance, leading to a wide acceptance of
  such a phenomenological Lagrangian. Furthermore, in such a theory,
  $\gjv$ should be identical for all currents (involving fields of a
  given charge) so as to preserve gauge invariance.

With the second pole in the photon propagator appearing only at
  $m_V^2 \, (1 + e^2 / \gv^2)$, the Lagrangians of
  eqns.(\ref{VMD_1}\&\ref{VMD_2}) are expected to give identical
  results for $q^2 \ll m_V^2$. Nonetheless, we will use the more
  common variant, namely eqn.(\ref{VMD_2}). This can be trivially
  extended to include multiple vector mesons. Similarly, $J^\mu$ can
  be generalised to different currents. Given this, the
amplitude for $\gamma^*\rightarrow \pi\rho$ can be expressed as
\begin{equation}
{\cal M}(\gamma^*\rightarrow \pi\rho)=\sum_{V=\varphi,\omega}\frac{e\, m_V^2}{\gv} \, 
      \frac{1}{q^2-m_V^2+im_V\Gamma_V} \, g_{V\pi\rho}\varepsilon_{\mu\nu\sigma\omega}p_1^\nu p_2^\sigma \epsilon^{*\omega}
\label{eqn:pi-rhoff}
\end{equation}
such that
\begin{equation}
 F_{\rm em}(q^2) = \sum_{V=\varphi,\omega}\frac{m_V^2}{\gv} \, 
      \frac{1}{q^2-m_V^2+im_V\Gamma_V} \, g_{V\pi\rho} \ .
\label{ff:pi-rhoff}
\end{equation}

In Table~\ref{table:coupling}, we list the processes used to estimate the values of the 
different coupling constants. In calculating these, we adopt the natural leading order form
for the three-point vector-vector-pseudoscalar meson vertex, {\em viz.}
$\dis \varepsilon_{\mu\nu\sigma\omega}p_{V1}^\nu p_{V2}^\sigma$.
\begin{table}[htb]
\centering
\begin{tabular}{|c|c|c|}
\hline
Coupling Constant & Value &Process \\
\hline
$g_{\omega_{782}}^\dagger$& 3.4$g_\rho$	  &	$\omega_{782}\rightarrow e^{-}e^{+}$\\
$g_{\omega_{782}\pi\rho}^\dagger$&	10.1$\gev^{-1}$&	$\omega_{782}\rightarrow\pi^+\pi^-\pi^\circ$\\
$g_{\phi}^\dagger	$&2.7$g_\rho$	& $\phi\rightarrow e^{-}e^{+}$	\\
$g_{\phi\pi\rho}$	&	0.2$\gev^{-1}$& $\phi\rightarrow\pi\omega_{782}$\\
$g_{\omega_{1420}}^\dagger$&0.9$g_\rho$&	$\omega_{1420}\rightarrow e^{-}e^{+}$ \quad \text{assuming Br. frac. = $10^{-5}$}\\
$g_{\omega_{1420}\pi\rho}$&	3.8$\gev^{-1}$& $\omega_{1420}\rightarrow\pi\rho$ \quad \text{assuming Br. frac. = 70\%}\\
\hline
\end{tabular}
\caption{\em Here, the value of $g_\rho = 5.01$ is calculated from the
decay constant of the $\rho$-meson while the decay rates of the given processes are taken from PDG\cite{PhysRevD.98.030001}. Note that the couplings marked by 
$\dagger$ are determined using vector meson dominance in the
respective processes.}
\label{table:coupling}
\end{table}

\begin{figure}
\centering
	  \subfigure[]{\label{fig:pi-rhoff}\includegraphics[width=75mm]{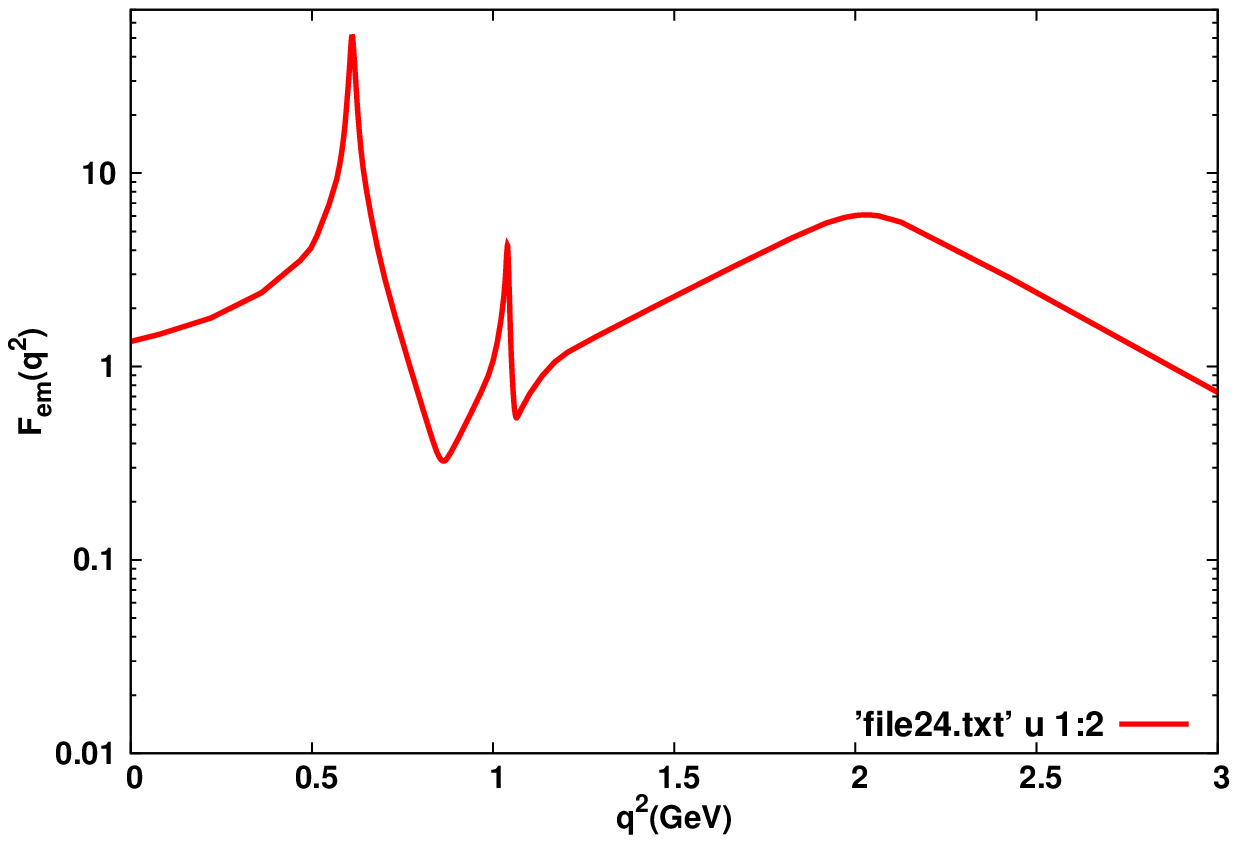}} 
	  \subfigure[]{\label{fig:relic_vector}\includegraphics[width=75mm]{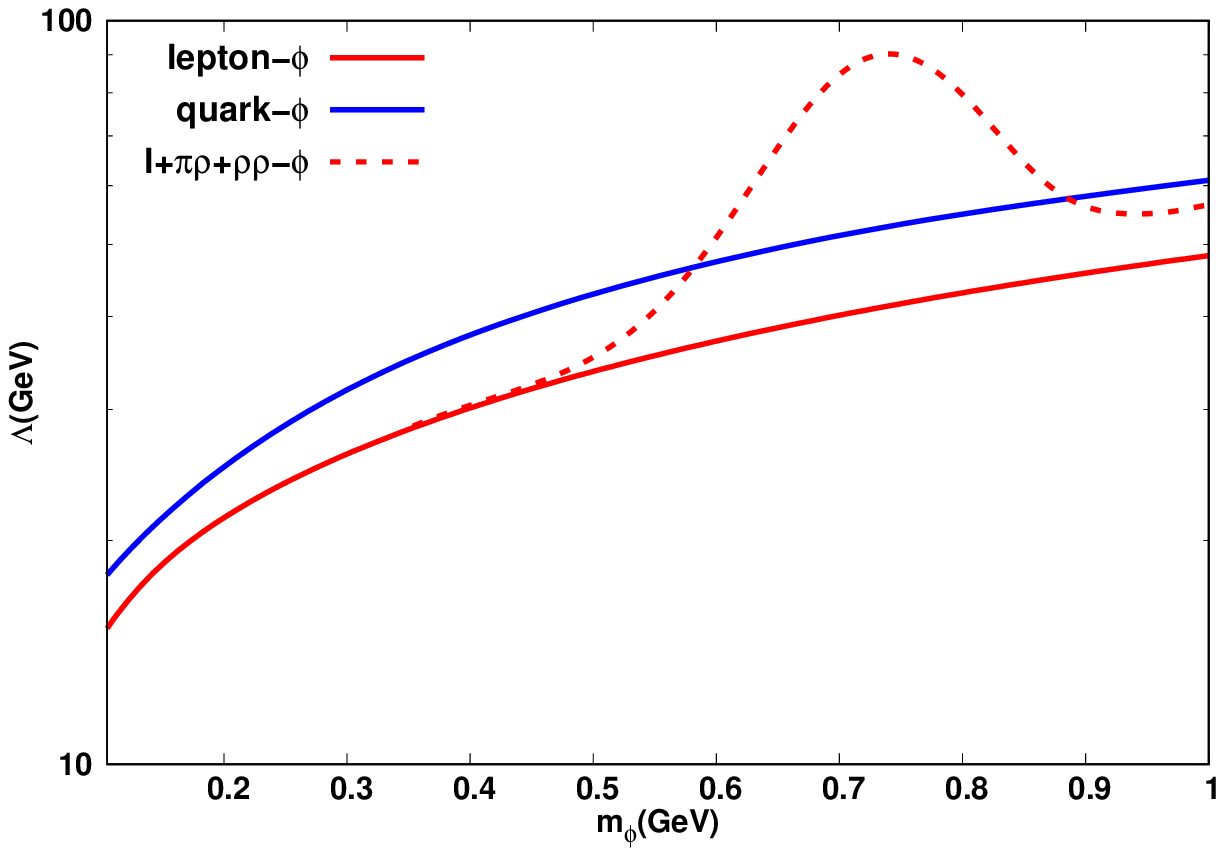}}
  \caption{\small \em{(a)The time like(TL) electromagnetic form factor estimated using the Vector Meson Dominance model.(b)Comparison of 
  contours in the $m_\varphi-\Lambda$ plane for the dimension-6 operators satisfying $ \Omega_{\varphi}h^2 = 0.1199 \pm 0.0022$ }.}
\end{figure}

Note that eqn.(\ref{ff:pi-rhoff}) gives a good approximation only for
relatively small range of $q^2$-values where the two poles dominate in
turn.  Consequently, we use this form only for the region 
$q^2 \leq 2.6\gev^2$. For large $q^2$ values ($q^2 \gapp
5 \gev^2$), perturbative results (incorporating $k_T$ factorization
{\em etc.})  are available and quite accurate (see
Ref.\cite{Hua:2018kho}). In between these two regions, we interpolate,
maintaining a $s^{-1}$ form (while poles do exist even in this region,
their contributions are small on account of the corresponding
vector-mesons having only suppressed couplings with a $\pi \rho$
pair). Finally, combining all the results, the electromagnetic form factor
for $\pi\rho$, as used by us, is shown in Fig.\ref{fig:pi-rhoff}. 
It behoves us to estimate the accuracy of our calculation of the form factor.
While comparison to data would be the natural check, unfortunately,
there exists no experimental data 
for the $q^2>0.5\gev^2$ region, thereby ruling out this possibility. For the 
$q^2 < 0.5 \gev^2$ region, data does exist, and a comparison with Ref.\cite{Maris:2002mz},
does show some deviation, but never exceeding 25\%. While even this 
may seem large, it should be realized that this particular range has very little
effect on the numerical results that we present next. Rather, this should be 
considered as the maximal theoretical possible uncertainty in our calculations.
Indeed, the errors are expected to be smaller for higher $q^2$ values, as 
the leading contributions have been accounted for.

\subsection{Revisiting Relic abundance for vector interactions}
We use these results to calculate the relic abundance and compare it
with the case of free quarks(Fig.\ref{r2}). The contour satisfying
$\Omega_\varphi h^2=0.1199$ is depicted in Fig.\ref{fig:relic_vector}.
Once again, we show the physically untenable curve corresponding to
the case of free quarks so as to facilitate easy comparison. Note
that, for large $m_\varphi$, the realistic curve approaches the
free-quark curve. This is quite similar to the case for the
dimension-5 operators and is reflective of the fact that, for such
$m_\varphi$ values, the free-quark approximation becomes better. More
importantly, there exists a large peak with a substantial width just
prior to the asymptotic region, where the free-quark approximation
would have grossly underestimated the annihilation cross-sections and,
hence, the sensitivity to the scale $\Lambda$. As can be realized from
the behaviour of the form factor as displayed in
Fig.\ref{fig:pi-rhoff}, this is but a reflection of the dominance of
the $\phi$ and $\omega$ mesons.  Despite the narrowness of the two
resonances, their closeness implies that, when convoluted with the
momentum spreads of the DM, they two peaks are no longer resolvable.
Rather, the two contributions add coherently, with the first one
dominating.

While the discussion above has concentrated only on the
  $\pi\rho$ final state, it is obvious that other final states too
  need to be taken into account for larger $m_{\varphi}$. Obvious
  candidates are states like $K K^*$, which dominate when the DM-pair
  couples to a strange-quark current.  Relatable by $SU(3)$ symmetry
  (albeit broken badly) to the $\pi\rho$ final state, we can use the
  same formalism for this case too. More interesting are final states
  comprising two vector mesons, such as $\rho\rho$. A very similar 
analysis would go through for this as well. Indeed, the break in the fall 
of the dashed curve in Fig.\ref{fig:relic_vector} and the subsequent rise 
for larger $m_\varphi$ owes its existence to the inclusion of the $\rho\rho$
state. The inclusion of even more states would drive this curve very close to 
 the blue curve. This is only to be expected as, for $m_{\varphi} > 2 \gev$, the
annihilation can be well-approximated by quasi-free quarks.

\section{CMB constraints}
\subsection{Effective relativistic degrees of freedom}
\label{sec:relativistic_dof}
 Energy injection from DM annihilation in the early universe can
 alter the effective number of relativistic degrees of freedom
$N_{\rm eff}$. Indeed, MeV-scale DM is especially constrained by these
observations. If the DM freezes out after the neutrinos have decoupled
(at $T = T_{\nu}^{\rm decoup}$), its annihilation will result in
heating the $e^-$--$\gamma$ plasma relative to the neutrinos, thereby
reducing the ratio of the neutrino and photon temperatures
($T_\nu/T_\gamma$). This results in a reduction of $N_{\rm eff}$ as
$N_{\rm eff}\propto(T_\nu/T_\gamma)^4$.  From standard cosmology
results, $N_{\rm eff}=3.046$~\cite{Ade:2015xua}, and only small
deviations from this value are allowed.

To find the expression for $T_f$ (equivalently, $x_f$), we can
approximate $\langle\sigma \,v\rangle\sim\sigma_0(1+b/x_f)$ (partial
wave expansion of the cross section).  On equating the interaction
rate ($\Gamma(x_f)$) with the expansion rate ($H(x_f)$), we get
\begin{equation}
x_f^{-1}=K~e^{-x_f}(1+b~x_f^{-1}) \qquad \text{where}  \qquad
 K = 0.038 \, g_\varphi \, g_\rho^{1/2} \, m_\varphi \, M_{pl} \, \sigma_0  \gg 1\ .
\end{equation}
Assuming that $Y \approx Y_{eq}^{\rm freeze-out}$ (or that
$n_\varphi(T_f)$ is equal to $n_\varphi$ today, {\it i.e.}, at $T_0=2.73~{\rm
  K}$), we can solve this iteratively. For example, at the first
iteration, the solution reads
\begin{equation}
x_f= \ln K+  \ln (\ln K) + \ln \left( 1 + \frac{b}{\ln K} \right) + \dots \ .
\label{x_f}
\end{equation}
This implies that $x_f$ depends, mainly, on $m_\varphi$, $g_\varphi$ and
$\sigma_0$. For a given particle, the magnitude of $\sigma_0$ is
similar for all the operators under consideration, provided the
respective Wilson coefficients are similar.

\begin{figure}[htb]
\centering
\includegraphics[width=75mm]{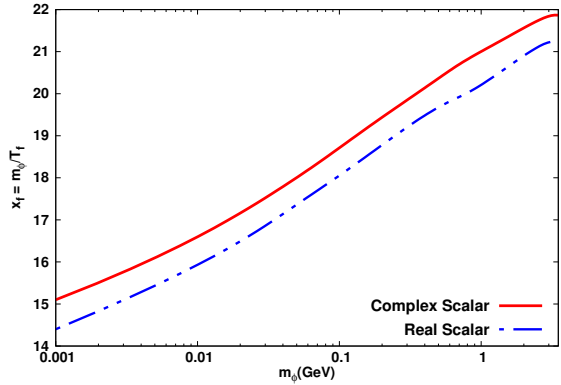}
  \caption{\em
  $x_f-m_{\varphi}$ curve satisfying $\Omega_{DM} h^2 = 0.1199 \pm 0.0022$ for the scalar operator. 
  For a particular scalar field, the corresponding curves  
  for the other operators are almost indistinguishable from those shown here.}
\label{fig:xf}
\end{figure}

In Fig.\ref{fig:xf}, we illustrate the decoupling temperature as a
function of $m_\varphi$ for the $\mathcal{O}^f_s$ operator.
With $\sigma_0$ being similar in magnitude for a given 
field, the value of $x_f$ would be very similar for
the other operators too. As the figure shows, for $m_{\varphi}>20 \mev$,
we have $T_f>$2.5 MeV. In other words, the scalar WIMP decouples prior
to neutrino decoupling ($T_\nu^{\rm decoup}=2$~MeV). Consequently, on
annihilation, it heats the neutrinos along with the photons and
electrons, preserving the standard result of $N_{\rm eff} \approx
3.046$.

The situation seemingly gets complicated for $m_\varphi \in [6,20]$~MeV,
when $T_f\approx T_{\nu}^{\rm decoup}$. However, note that the bulk of
the entropy transfer due to DM annihilation still occurs at $T\sim
m_\varphi/3$, {\em i.e.}, prior to $\nu-$decoupling, and, hence, the
model is safe from such constraints. Thus, for a complex scalar field
$\varphi$, it is the $m_{\varphi}<$6~MeV range which is constrained by
$N_{\rm eff}$.  Similarly, for a real scalar, the limit is 3
MeV. These results are in consonance with those in
Ref.\cite{Ho:2012ug}. However, it should be reiterated that this is
operative only in (standard) scenarios wherein the DM is presumed to
have been produced thermally and having been in thermal equilibrium
with the SM particles\footnote{If light DM enters thermal equilibrium
  with the SM after neutrino-photon decoupling, then the constraints
  from measurements of $N_{\rm eff}$ are significantly
  relaxed\cite{Berlin:2017ftj}.}, for a long enough phase with its
abundance being defined thereby.  For a non-thermal DM, the couplings to
neutrinos and/or photons will have to be tuned such that the model
satisfies above constraints.

\subsection{CMB observations and Indirect Detection}
 The Cosmic Microwave Background Radiation encodes information about
 the thermal history of the early universe, and is well described by
 SM physics. On the other hand, DM annihilation, at early times, into
 high energy photons or charged particles can not only heat the gas,
 but can also lead to atomic excitations and even its ionization.
 This increase in the amount of the ionized fraction
 causes an increase in the width of the last scattering surface,
 thereby affecting the power spectrum of the
 CMB~\cite{Chen:2003gz,Madhavacheril:2013cna}.

The energy injected by DM annihilation into the CMBR depends on its
number density $n_{\varphi}$ at that epoch, the rate of its annihilation
into (charged) SM particles and the nature of the cascade of particles
produced after annihilation. Due to this cascading, not the entire
energy is transferred to the CMB (or the plasma in equilibrium with
it) but only a fraction. To calculate the amount of the energy
transferred, one needs to track the evolution of the hydrogen and
helium ion fractions, and the spectra of $e^\pm$ and photons at that
epoch. With these being temperature-- and, hence,
redshift--dependent, we are faced with a redshift dependent efficiency
function $f(z)$ that describes the fraction of the energy absorbed by
the CMB plasma.  It has been argued~\cite{Slatyer:2015jla}, though,
that the effect of $f(z)$ can be well-approximated by an effective,
but redshift independent, efficiency function $f_{\rm eff}$. Indeed,
Ref.\cite{2012PhRvD..85d3522F} demonstrated that, given a set of
$f(z)$ functions for a WIMP, the impact of an appropriately chosen
$f_{\rm eff}$, on the CMB, is identical at the sub-percent level. This
is the simplification that we shall adopt.

The rate of energy deposited, into the CMB, by
  DM pair annihilation per unit time per unit volume is given by:
\[	
\frac{dE}{dt\,dV}=\rho_c^2 \, \Omega_\varphi^2 \, (1+z)^6 \, P_{\rm ann}(z) \ ,
\]
where $z$ is the redshift of the epoch, and $\rho_c$($\Omega_{\varphi}$)
is the critical density of the universe (DM relic abundance) today,
{\it i.e.}, at $z= 0$. The factor $(1+z)^6$ just encapsulates the
standard evolution of the dark matter number density (note that the
annihilation rate is proportional to $n_\varphi^2$). CMB
observations\cite{Ade:2015xua} constrain $P_{\rm ann}<4.1\times
10^{-28}$cm$^3$s$^{-1}$GeV$^{-1}$.  To translate this into the allowed
region in the $m_\varphi-\Lambda$ plane, we need $f(z)$ (or,
equivalently, $f_{\rm eff}$) and $\left<\sigma v\right>$. Here, $v$ is
the relative velocity (in units of the velocity of light)
of the second DM particle in the rest frame of
the first. For a complex scalar field, the thermal average of the
annihilation cross sections for different operators are given by
\begin{eqnarray}
  \left<\sigma v\right>_{\mathcal{O}^f_{s}} &\simeq& \frac{1}{4\pi\Lambda^2} \sum\limits_f
\sqrt{1-\frac{m_f^2}{m_{\varphi}^2}} \left[\left(1-\frac{m_f^2}{m_{\varphi}^2}\right)
+ \frac{1}{8}\left(-2+5\frac{m_f^2}{m_{\varphi}^2}\right) \left<v^2\right>\right] \\
\left<\sigma v\right>_{\mathcal{O}^f_{p}} &\simeq& \frac{1}{4\pi\Lambda^2}  \sum\limits_f
\sqrt{1-\frac{m_f^2}{m_{\varphi}^2}} \left(2+\frac{-2 + 3 (m_f^2/m_{\varphi}^2)}{8(1-(m_f^2/m_{\varphi}^2))}\left<v^2\right>\right),\\
\left<\sigma v \right>_{\mathcal{O}^f_{v}} &\simeq& \frac{1}{12\pi\Lambda^4}  \sum\limits_f
\sqrt{1-\frac{m_f^2}{m_\varphi^2}} m_\varphi^2 \left(2+\frac{m_f^2}{m_\varphi^2}
\right)\left<v^2\right>,\\
\left<\sigma v\right>_{\mathcal{O}^f_{a}} &\simeq& \frac{1}{6\pi\Lambda^4} \sum\limits_f
\left(1-\frac{m_f^2}{m_\varphi^2}\right)^{3/2} m_\varphi^2 \left<v^2\right>,\\
\left<\sigma v\right>_{\mathcal{O}_{\gamma,\tilde{\gamma}}} &\simeq& 
\frac{2}{\pi\Lambda^4} 
 m_\varphi^2 \left[1-\left(\frac{9}{16}\left<v^4\right>\right)\right] \ .
\end{eqnarray}
 For a real scalar field, similar expressions hold, but with an extra
 factor of $4$. As derived in appendix A, we have
 $\left<v^2\right>=6~T/m_{\varphi}$ and $\left<v^4\right>=60
 \,T^2/m_{\varphi}^2$.
 
 The dependence of $\langle \sigma v\rangle$ on
  $\langle v^n\rangle$ is easy to understand in terms of the angular
  momenta, especially if the DM-pair is viewed as a composite
  pseudoparticle (with some angular momentum) decaying into a
  SM-pair. For  $\mathcal{O}^f_{s}$ and
$\mathcal{O}^f_{p}$, the
  amplitude has both s-wave and p-wave components, whereas for
$\mathcal{O}^f_{v}$ and $\mathcal{O}^f_{a}$, no s-wave component can exist on account of the inherent angular
  momentum of the initial state.
  Similarly, for $\mathcal{O}_{\gamma}$ and
  $\mathcal{O}_{\tilde{\gamma}}$, no p-wave component may exist as it
  would require the diphoton state to exist in an antisymmetric
  state.

With the DM being nonrelativistic, the CMB constraints for
p-wave annihilation are weaker compared to those for the 
  cases driven by s-wave DM annihilation.  
For pure s-wave annihilation, $\left<\sigma v\right>$ is
independent of velocity and hence $P_{\rm ann}$ is a
redshift-independent parameter. For vector couplings, 
on the other hand, we
need to estimate
 the velocity of DM at the epoch where these interactions are
significant. 
To this end, we consider the epoch of kinetic decoupling. 
When the Hubble rate equates
the rate of scattering, the dark matter can no longer 
reach kinetic equilibrium with the plasma through 
a high momentum exchange rate; the WIMPs kinetically decouple from 
the plasma and attain free-streaming. This allows us to write
\[
P_{\rm ann}^{\text{p-wave}}=f_{\rm eff} \, \frac{\langle\sigma v\rangle_{\text{CMB}}}
                                        {m_{\varphi}}
\]
and use 
\[
\left<v^2\right>_{\text{CMB}}=\left<v^2\right>_{kd}\frac{(1+z_{CMB})^2}{(1+z_{kd})^2}=\left<v^2\right>_{kd}\frac{T^{\varphi}_{\text{CMB}}}{T^{\varphi}_{kd}}
\]
to calculate $\langle\sigma v\rangle_{\text{CMB}}$.  As the DM was,
hitherto, in kinetic equilibrium with the plasma,
$\left<v^2\right>_{kd}=\sqrt{6 \, T_{kd}/m_{\varphi}}$.  A conservative
estimate gives\footnote{The exact value for different operators vary
  from $10^{-4}$ to $10^{-6}$, as can be estimated by equating the
  rate for elastic scattering (DM and SM) to the expansion rate, \it
  i.e., $ n_{rel} \, \Gamma_{\rm elastic}\sim H(T_d)$.}
$x_{kd}=T_{kd}/m_\varphi\sim10^{-4}$.  Using $z_{CMB}\sim 1100$,
we may estimate $T^{\gamma}_{CMB}=T^{\gamma}_{\rm today}(1+z_{CMB})$.
The temperature of $\varphi$ at recombination depends on the kinetic
decoupling temperature. As long as the DM remains kinetically coupled
to the plasma, we have $T^\gamma_{kd}=T^{\varphi}_{kd}$. Once the DM 
decouples kinetically, its temperature at a redshift $z\sim
z_{CMB}$ is decided by its non-relativistic nature, i.e,
\[
\dis\frac{T^{\varphi}_{CMB}}{T^\varphi_{kd}}=\dis\Big(\frac{z_{CMB}}{z_{kd}}\Big)^2=\dis\Big(\frac{T^\gamma_{CMB}}{T^\gamma_{kd}}\Big)^2 \ .
\]
This implies
  \[
\left<v^2\right>_{\text{CMB}}=\frac{(T^\gamma_{CMB})^2}{m_\varphi\,T^\varphi_{kd}}=\frac{(T^\gamma_{CMB})^2}{m_\varphi^2}x_{kd} \ ,
\]
or, in other words, 
$\left<v^2\right>_{\text{CMB}}$ of $\varphi$ of
$m=1\mev$ exceeds that for $m=1\gev$ 
by a factor of $10^9$.

As for the effective efficiency factor $f_{\rm eff}$, it depends upon the details
of the model (in our case, the relative sizes of the Wilson
coefficients), and rather than calculate it explicitly, we allow it to
vary within the range $0.4 < f_{\rm eff} < 1$ which is commensurate
with that advocated in Ref.\cite{Gonzalez-Morales:2017jkx}. As
we shall see later, our results are not going to be very sensitive to
the exact choice.

\begin{figure*}[htb]
\centering
	  \subfigure[]{\label{c1}\includegraphics[width=75mm]{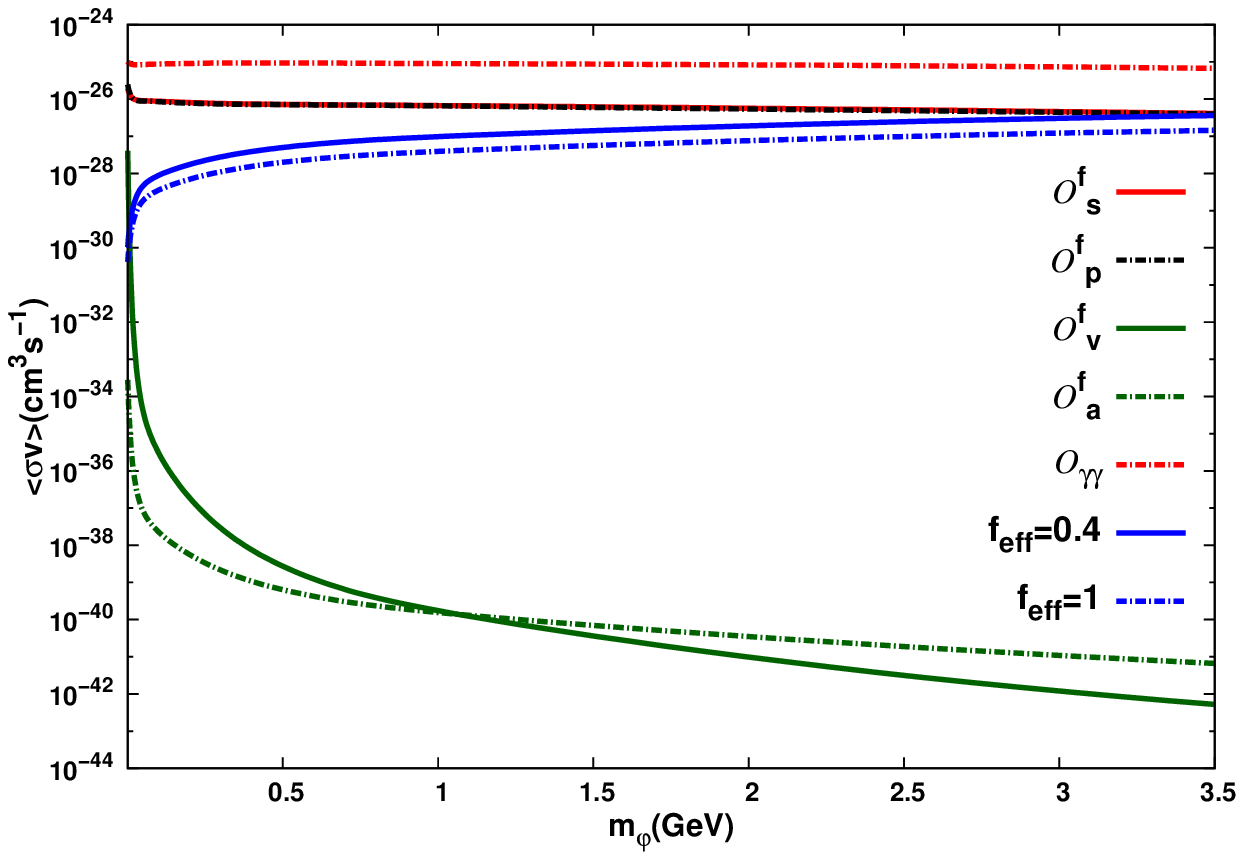}} 
	  \subfigure[]{\label{c2}\includegraphics[width=75mm]{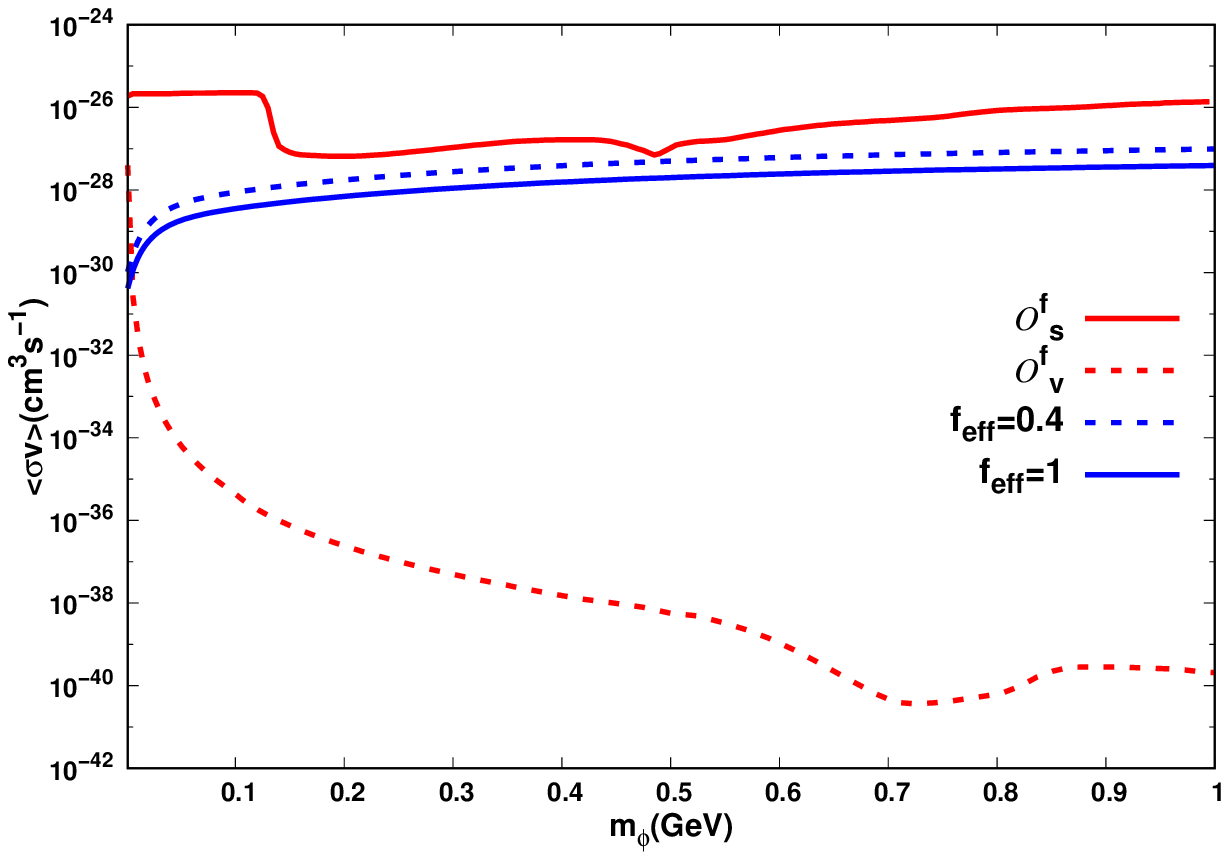}}
  \caption{\em We present the $\langle\sigma
    v\rangle(\varphi\varphi\rightarrow e^-e^+)-m_\varphi$ plane obtained using
    those value of $\Lambda$ that satisfy $\Omega_\varphi
    h^2=0.1199\pm0.0022$ for the case when DM is allowed to (a)
    annihilate into leptons and free quarks and (b) annihilate into
    leptons and bound states.}
\label{fig:cmb} 
\end{figure*}

Since DM annihilation to electrons and
photons give us the tightest constraints, in Fig.~\ref{fig:cmb}, we depict the
value of $\left< \sigma v\right>(\varphi\varphi\rightarrow e^-e^+)$, as a
function of $m_\varphi$ for the respective operators of
eq.(\ref{the_operators}) and different final states. In Fig.~\ref{fig:cmb}, we depict the value of
$\left< \sigma v\right>(\varphi\varphi\rightarrow e^-e^+)$, as a function of
$m_\varphi$ for the respective operators of eq.(\ref{the_operators}) and
different final states. In each case, $\Lambda$ is chosen to be the
maximum allowed for by the measurement of the relic density, {\it
  viz.} the condition $\Omega_\varphi \leq \Omega_{DM}$. Thus, it is the
area {\em above} a curve that is allowed. Also depicted, in solid
blue, is the curve corresponding to the aforementioned CMB observation
{\it viz.}  $P_{\rm ann}\sim 4.1\times
10^{-28}$cm$^3$s$^{-1}$GeV$^{-1}$. The top (bottom) curves correspond
to $f_{\rm eff} = 0.4 \, (1)$ respectively. Clearly, it is the area
{\em below} these curves that is allowed. Had the quarks in the final state
been truly free, we would, thus,
have faced a seeming disagreement between
the two sets of observations, at least for smaller values of
$m_\varphi$. However, before we entirely discard such a mass range, 
we need to reconsider the correction wrought by 
considering bound states instead. As Fig.\ref{c2} shows, this reduces 
the disagreement to a very large extent. The inclusion of even more 
bound states in the calculation of relic abundance, would 
    have further reduced the remaining disagreement 
    (especially for $m_\varphi \gapp 500\mev$). To appreciate this,
    recognize that such an inclusion would raise the DM annihilation cross section (into hadronic states) even further, 
thereby implying a raise in the preferred value of $\Lambda$ and, hence, a 
suppression in the $\varphi\varphi \to e^+ e^-$ rates.

In addition, it should be borne in mind that the said disagreement
depends not only upon our understanding of the early universe being
perfect, but also on several key assumptions. For example, consider
the case of the non-thermal DM,
where the annihilation cross sections are very small and the final
relic abundance is completely determined by its initial abundance
which, in turn, depends on the model at hand. For such small cross
sections, these limits can be evaded easily. For scenarios that fall
somewhere in between the non-thermal and rigorously thermal DM, the
constraints would need to be scaled appropriately. 

\section{Summary and Conclusions}
In this work, we have systematically investigated the interactions of
a light scalar DM particle with the SM sector within the framework of
an effective field theory. This entails examining the constraints
imposed by astrophysical/cosmological observations such
as the CMB power spectrum, the deduced value of the relic abundance as
well as the comparison of the photon and neutrino temperatures. 

In the first part of the work, we considered leptons and free quarks
as final states (while this is not a good approximation given
  the smallness of the DM mass and, hence, energies available, it
  serves to illustrate certain features and sets the stage for the
  second part of the study) and analysed the
constraints imposed by the cosmologically deduced value of the relic
density ($\Omega_{\rm DM}\, h^2$). This allowed us to obtain relatively
robust upper limits on the
scale $\Lambda$ of the effective field theory; as a higher
value for $\Lambda$ would imply a smaller DM
annihilation cross section, and, hence, a larger abundance. For the
dimension-6 operators under consideration, we have $\Lambda \propto
\sqrt{m_\varphi}$ so as to produce the right abundance, while, for the
dimension-5 operators, $\Lambda \sim 10^4 \gev$ with only a weak
dependence on $m_\varphi$. The value of $\Lambda_{\rm max}$ corresponding to the
  dimension-6 operators might, at the first sight, seem too low to
  have escaped detection in terrestrial experiments.  However, this is
  not quite so as we would demonstrate in the accompanying
  paper\cite{paper_2}.

As already stated, for DM masses below a couple of GeVs, it is not a
good approximation to assume that a pair of DM particles may annihilate
into a pair of free quarks as DM of such a mass would freeze out only
around the QCD phase transition temperature, leaving them with very
little overall energy. Consequently, we should, instead, re frame the
analysis in terms of bound states. Given that baryon production is suppressed, we concentrate on
mesons, devising appropriate methodology to determine the leading
annihilation cross sections into such states. Chiral perturbation
theory as well as techniques of dispersion analysis are used to
obtain the effective couplings of DM with a host of light mesons, not
only the pseudoscalars such as pions and kaons, but vectors as
well. However, it is the pseudoscalar states that dominate for
dimension-5 operators and obtaining the right relic abundance would need
the scale of the effective theory to be related to the DM mass as
$\Lambda \propto F(q^2)m_\varphi^{-2}$. For dimension-6 operators, on the
other hand, annihilation to final state (pseudo)scalar mesons is
$v^4$ suppressed and, hence, we must include the vector mesons in the
mix. For a final state comprised of a pair of vector mesons, very good
results can be obtained using the analogues of the time-like
electromagnetic form-factors. For a pseudo scalar-vector combination,
on the other hand, a combination of data and the vector-meson-dominance model
does the job. This leads to $\Lambda \propto \sqrt{m_\varphi\,F(q^2)}$ for vector
interactions. What is particularly heartening to see is that the
inclusion of progressively more states brings the results closer and
closer to that obtained with free quarks. This lends credence to the
belief that the results found herein present a very good approximation
and can be made even more robust by the inclusion of just a few more
states at best.

 An orthogonal constraint emanates from the requirement that the
annihilation of the DM does not significantly alter the ratio of the
neutrino and photon temperatures, an observable often recast as
$N_{\rm eff}$ (or, the effective number of neutrino-like species).
For the effective Lagrangians under consideration, once the
requirement of reproducing the right relic abundance is imposed, the
freeze-out temperature $T_f \sim d_1 \, m_\varphi \, \exp(m_\varphi /
d_2)$ with the constants $d_{1,2}$ being only very weakly dependent
on the exact nature of the current-current structure. A consequence
is that the constraints are strong only for $m_\varphi \lapp 6 \mev$,
as for higher values of $m_\varphi$, the bulk of the entropy transfer
to the plasma, takes place {\em before} $T \approx m_\varphi /3$, and,
hence, before the neutrinos have decoupled. Even this constraint
(for $m_\varphi \lapp 6 \mev$) can be evaded if the DM had a
non-thermal origin. However, with the dynamics of such DM being very
sensitive to the spectrum and the structure of the theory, it does
not easily lend itself to the effective Lagrangian treatment.

A competing constraint emanates from the shape of the CMB spectrum.
The lack of significant distortions in the same puts an upper limit on
the rate of DM annihilation ($P_{\rm ann}$) to, for example, $e^\pm$
or photon-pairs. The lower limit on $\Lambda$ that this translates to
is, often, in ostensible contradiction with the aforementioned
values of $\Lambda_{\rm max}$. These two opposing constraints, thus,
seemingly rule out such a light DM (within the ambit of an effective
theory). However, with the inclusion of bound states, the disagreement
between the observables $\Omega_\varphi h^2$ and $P_{\rm ann}$ is rendered
comparatively small. It should be realized, though, that existence of
even such a small ``discrepancy'' depends crucially on the assumption
that, in the early universe, the DM was in exact thermal equilibrium
with the SM sector. If this restriction is released, or, in other words,
a non-thermal initial condition on the DM allowed for, the constraints
from $P_{\rm ann}$ are eased sufficiently enough to permit a large overlap
with the parameter space allowed by $\Omega_{\rm DM}$. 

Similar to the CMB constraints, the inclusion of bound states slightly changes 
the interpretation of the results in the direct detection experiments which
 we have discussed in the companion paper.

In summary, we may conclude that a substantial fraction of the parameter
space of light scalar DM is still viable. Furthermore, an accurate estimation
of the cosmological constraints needs the proper inclusion of bound states. 
While the inclusion of the few light mesons already given us a fast converging 
and robust result, the remaining uncertainties can be reduced in a straightforward 
(though painstaking) manner by the inclusion of even more states.

\appendix
\section{Appendix A}
For any observable $f(p_1, p_2, \dots, p_n)$, 
constructed of the momenta of $n$ particles 
of a gas at equilibrium,
the thermal average is given by
\[
\langle f \rangle = \frac{\dis\int \frac{d^3 p_1}{(2 \, \pi)^3} \cdots 
                           \int \frac{d^3 p_n}{(2 \, \pi)^3} \, 
                           e^{\left(- \sum_i E_i / k_B T \right)} \,
                           f(p_1, p_2, \dots, p_n)}
                 {\dis\int \frac{d^3 p_1}{(2 \, \pi)^3} \cdots 
                           \int \frac{d^3 p_n}{(2 \, \pi)^3} \, 
                           e^{\left(- \sum_i E_i / k_B T \right) }}  \ ,
\]
where $E_i$ is the energy of the $i$'th particle. For a single non-relativistic
species, 
\[
E_i \simeq m + \frac{p_i^2}{2 m} \simeq m+ \frac{m \, v_i^2}{ 2} \ .
\] 
Defining
the relative velocity between two particles as 
\begin{equation} 
v_{\rm rel} \equiv \vec v_1 - \vec v_2 \ ,
\end{equation} 
we have
\begin{equation*}
\langle v_{\rm rel}^2\rangle =
\frac{
   \dis \int d^3v_1 \, d^3v_2 \, 
            \exp\left( \frac{-m \, (v_1^2 + v_2^2)}{2 \, k_B \, T} \right)
                 |\vec v_1 - \vec v_2|^2 
     }{
   \dis \int d^3v_1 \, d^3v_2 \, 
            \exp\left( \frac{-m \, (v_1^2 + v_2^2)}{2 \, k_B \, T} \right)
     }
\end{equation*}
Henceforth, we choose $k_B = 1$. Effecting a change of variables, namely
\[
(\vec v_1, \vec v_2) \to 
  \left(\vec v_{\rm rel}, \vec v_{\rm cm} \equiv \frac{\vec v_1 + \vec v_2}{2} 
  \right) \ ,
\]
the Jacobian is unity, and 
\begin{equation}
\barr{rcl}
\dis \langle v_{\rm rel}^2\rangle 
&= & \dis 
   \frac{\int d^3 v_{\rm cm} \, d^3 v_{\rm rel} \, v_{\rm rel}^2 \,
       \exp\left[ (- m/T) \left(v_{\rm cm}^2 + v_{\rm rel}^2 / 4 \right) \right]}
       {\int d^3 v_{\rm cm} \, d^3 v_{\rm rel} \, 
       \exp\left[ (- m/T) \left(v_{\rm cm}^2 + v_{\rm rel}^2 / 4 \right) \right]}
\\[3.5ex]
&= & \dis 
   \frac{\int d v_{\rm rel} \, v_{\rm rel}^4 \,
       \exp\left[ - m \, v_{\rm rel}^2 / 4 T \right]}
{\int d v_{\rm rel} \, v_{\rm rel}^2 \,
       \exp\left[ - m \, v_{\rm rel}^2 / 4 T \right]}
\\[3.5ex]
& = & \dis \frac{6 \, T}{m} \ .
\earr
\end{equation}

Similarly,
\begin{equation}
\barr{rcl}
\dis \langle v_{\rm rel}^4\rangle 
&= & \dis 
   \frac{\int d v_{\rm rel} \, v_{\rm rel}^6 \,
       \exp\left[ - m \, v_{\rm rel}^2 / 4 T \right]}
{\int d v_{\rm rel} \, v_{\rm rel}^2 \,
       \exp\left[ - m \, v_{\rm rel}^2 / 4 T \right]}
= \frac{60 \, T^2}{m^2} \ .
\earr
\end{equation}

\bibliographystyle{JHEP}
\bibliography{reference}

\end{document}